\documentclass[article,12pt]{elsarticle}
\usepackage[a4paper,text={400pt,650pt},centering]{geometry}
\usepackage[utf8]{inputenc}
\usepackage{amssymb,amsmath}
\usepackage{relsize}
\usepackage{verbatim}
\usepackage{appendix}
\usepackage{mathtools}
\usepackage{tikz}
\usetikzlibrary{decorations.markings}
\usepackage{braket}
\usepackage[new]{old-arrows}
\usepackage{mathtools}

\usepackage{bbold}
\usepackage{tablefootnote}
\usepackage{float}
\usepackage{amsmath,bm}
\usepackage{enumerate}
\usepackage{pgfplots}
\usepackage{fancyhdr}
\usepgfplotslibrary{fillbetween}
\usetikzlibrary{patterns}
\pgfplotsset{compat=1.12}
\usepackage{mathtools}
\usepackage{graphicx}
\usepackage{caption}
\usepackage{subcaption}

\usepackage{graphicx}
\usepackage{epstopdf}
\usepackage{fancyhdr}

\usepackage[compat=1.0.0]{tikz-feynman}
\usepackage{amsfonts,amssymb,amsthm,epsfig,epstopdf,url,array}
\def\ov{\overline}
\usepackage{braket}

\usepackage{enumerate}
\usepackage{caption}

\usepackage{lineno,hyperref}
\modulolinenumbers[5]

\journal{A}
\biboptions{sort&compress}   

\let\today\relax
\makeatletter
\def\ps@pprintTitle{%
    \let\@oddhead\@empty
    \let\@evenhead\@empty
    \def\@oddfoot{\footnotesize\itshape
         {\sc \begin{tabular}{l}
                MPP--2024--32 \\MITP--24--035  
                \end{tabular}} \hfill\today}%
    \let\@evenfoot\@oddfoot
    }
\makeatother

\usepackage[english]{babel}

\theoremstyle{remark}

\usepackage[shortcuts]{extdash} 
\allowdisplaybreaks

\newcommand{\sect}[1]{ \section{#1} \setcounter{equation}{0} }

\newcommand{\req}[1]{(\ref{#1})}
\def\vev#1{\langle #1 \rangle}

\def\bet{\beta}
\def\fc#1#2{\frac{#1}{#2}}
\def\h{\frac{1}{2}}
\newcommand{\nwc}{\newcommand}
\nwc{\ba}  {\begin{array}}
\nwc{\ea}  {\end{array}}
\nwc{\bdm} {\begin{displaymath}}
\nwc{\edm} {\end{displaymath}}

\nwc{\bea} {\begin{equation}\ba{lcl}}
\nwc{\eea} {\ea\end{equation}}

\nwc{\be} {\begin{equation}}
\nwc{\ee} {\end{equation}}

\nwc{\bda} {\bdm\ba{lcl}}
\nwc{\eda} {\ea\edm}

\nwc{\bc}  {\begin{center}}
\nwc{\ec}  {\end{center}}

\nwc{\ds}  {\displaystyle}

\nwc{\nn} {\nonumber}
\nwc{\nnn} {\nonumber \vspace{.2cm} \\ }
\nwc{\ra}{\rightarrow}
\nwc{\lra}{\longrightarrow}
\def\lf{\left}\def\ri{\right}

\nwc{\p} {\partial}
\def\IR{{\bf R}}
\def\Fc{{\cal F}}
\def\ap{\alpha'}
\def\lng{\langle}
\def\rng{\rangle}
\def\Mc{{\cal M}}

\def\ov{\overline}

\def\al{\alpha}
\def\bet{\beta}

\def\IR{{\bf R}}
\def\IC{{\bf C}}
\def\IP{{\bf P}}
\def\IZ{{\bf Z}}

\def\Lc{{\cal L}}
\def\Fc{{\cal F}}

\def\al{\alpha}

\def\si{\sigma}
\def\om{\omega}
\def\bet{\beta}\def\bet{\beta}

\begin{document}
\begin{frontmatter}

\title{{\sc  One--loop Double Copy Relation \\[2mm]
from Twisted (Co)homology\vskip1.5cm}}

\author{Pouria Mazloumi$^{a}$\ and\  Stephan Stieberger$^{b,c}$}


\address{$\ $\\[1mm]
$^a$Johannes Gutenberg-Universit\"at Mainz, \\55099 Mainz, Germany  \\[2mm]
$^b$Max-Planck Institut f\"ur Physik\\Werner--Heisenberg--Institut, 
85748 Garching, Germany\\[2mm]
$^c$Kavli Institute for Theoretical Physics, Santa Barbara, CA 93106, USA 
$\ $\\[1mm]}

\address{{\it Emails:} pmazloumi@uni-mainz.de, stieberg@mpp.mpg.de \vskip2cm}

\begin{abstract}

We propose a geometric relation between closed and open string amplitudes at one-loop. 
After imposing a homological splitting on the world--sheet torus, twisted intersection theory is used to establish a one--loop double copy relation. The latter expresses 
a closed string amplitude by  a pair of open string amplitudes and twisted intersection numbers.
These inner products on the vector space of twisted differential forms are related to the twisted homology and cohomology groups associated with the Riemann--Wirtinger integral.

\end{abstract}


\end{frontmatter}

\newpage

\setcounter{tocdepth}{2}
\tableofcontents
\newpage

\section{Introduction}

A geometric description of scattering amplitudes is appealing because it incorporates concepts like locality, unitarity and symmetries more naturally than in the conceptual Feynman diagram approach, where these properties  are generically imposed by hand.
This fact leads to a new understanding of computing amplitudes from first principles.
Certainly, a prime example is  the geometry of a Riemann surface with genus $g$ as string world--sheet describing the interactions of  strings at $g$--loop order. In this case all Feynman diagrams simply follow from considering certain limits of the underlying Riemann surface.
In fact, the geometry of the moduli space of compact Riemann manifolds with punctures is ubiquitous for a geometrical description of  scattering amplitudes of particles. This geometry appears for computing intersection numbers of twisted differential forms, for formulating amplitudes in terms of  the Cachazo--He--Yuan (CHY) description and as world--sheet for closed string amplitudes. The latter are formulated as multi--dimensional complex integrals over the Riemann manifolds, which in turn can be described by twisted de Rham cohomology.
Twisted de Rham cohomology  is suitable for considering multi--valued differential forms because they naturally appear on the string world--sheet.
In this context, twisted intersection theory provides a geometric framework to describe the underlying structure of amplitudes by intersection numbers. The latter are inner products on the vector space of twisted differential forms defined by the twisted covariant derivative.

The famous Kawai--Lewellen--Tye (KLT) relations express a tree--level closed string amplitude as a weighted sum over squares of tree--level open string amplitudes \cite{klt}.
Since the lowest mode of the closed superstring is a graviton and that of the open superstring a gluon, the aforementioned relation gives rise to a gauge/gravity correspondence linking gravity and gauge amplitudes at the perturbative tree--level. 
The relation between gauge and gravity amplitudes was then reformulated as double--copy structure by Bern, Carrasco and Johansson (BCJ) based on a duality between color and kinematics derived from the (conjectural) existence of relations between partial gluon subamplitudes \cite{BCJ1}. The general proof of such tree--level BCJ amplitude identities was subsequently presented by using string theory and the power of world--sheet monodromy properties \cite{Stie,Bjerrum-Bohr:2009ulz}. The formulation of the color--kinematics duality has since then seen a large number of further extensions, we refer to \cite{Bern:2019prr} for a comprehensive review. Only recently an one--loop analog of the KLT relations  have  been derived at one--loop string theory level \cite{Stieberger:2022lss}. Furthermore, in \cite{Stieberger:2023nol} the underlying double copy structure has been investigated.

KLT like relations relate gravity amplitudes to squares of gauge amplitudes involving an intersection matrix (or KLT kernel) \cite{Bern:1998sv,Bjerrum-Bohr:2010pnr}. The latter is interesting on its own and constructing it by first principles has been received a lot of interests recently, cf. \cite{Chi:2021mio}. In addition, the opposite question has been addressed how a (single--valued) amplitude can be written as a pair of amplitudes with monodromies \cite{Baune:2023uut}. Furthermore, at tree--level the KLT kernel has a geometric interpretation in terms of twisted intersection theory \cite{Mizera:2017cqs}.
Likewise, tree--level double copy structures of amplitudes have a natural formulation in terms of
twisted intersection numbers. In fact, a rich catalogue of twisted forms has been constructed recently to formulate new double copies \cite{M1,MS,Mazloumi:2022nvi}.

The one--loop KLT results of \cite{Stieberger:2022lss,Stieberger:2023nol} derived from contour deformation on the genus--one torus should have an interpretation in terms of twisted (co)homology.
So far, not much is known at one--loop by means of formulating amplitudes by twisted differentials on the elliptic curve. Albeit, at the mathematical level there are some important works dealing with Riemann--Wirtinger integrals first introduced by Mano \cite{Mano,Mano2} and then further discussed in \cite{Watanabe,goto2023intersection,ghazouani2016moduli} with a relation between the  Riemann--Wirtinger integral and the Felder--Varchenko integral solution of the KZB equation in \cite{Watanabe}.

In this work we establish a one--loop double copy relation expressing
a closed string amplitude by  a pair of open string amplitudes and twisted intersection numbers.
The latter  are related to the twisted homology and cohomology groups associated with the Riemann--Wirtinger integral. From the mathematical side this relation establishes a twisted Riemann period relation at genus one.
For our setup we need to impose $\Re(\tau)=0$ in order to establish a factorization between holomorphic and anti--holomorphic sectors and perform multi--dimensional complex torus integrations. 
On the other hand, a complementary work has recently appeared for the {\it single} complex integration at generic complex structure \cite{Bhardwaj:2023vvm}. In this case, in order to match to a double copy of  doubly periodic Riemann--Wirtiniger systems  a constraint needs to be imposed relating complex structure, positions and  kinematic invariants. At present it is not  clear how this constraint influences the remaining position integrations. We shall further comment on this in Section~\ref{Conclusion}.

\sect{Twisted de Rham theory and intersection numbers of differential forms}

Twisted de Rham theory with a covariant derivative
\be
\nabla_\om=d+\omega\wedge\ ,
\ee
involving an exterior differentiation $d$ on a complex manifold $X$ and a holomorphic closed one--form $\om$ on $X$, has been pioneered by Aomoto, Deligne, Gelfand and Kita, cf.~\cite{book}.
Integrals of multi--valued functions are formulated as pairings between integration cycles $\Delta$ and corresponding cocycles  as their integrands or likewise pairings between twisted homology and twisted cohomology  class
\be\label{PAIR}
(\Delta\otimes KN,\varphi)=\int_\Delta KN\; \varphi\ ,
\ee
with a $p$--dimensional oriented smooth simplex $\Delta$ in $X$, $KN$ a multi--valued solution to $\nabla_{-\omega} KN=0$\footnote{Stokes' theorem yields $KN^{-1}\;d (KN \varphi)=\nabla_{\omega} \varphi=0,$ with $\omega=d \ln(KN)$. Therefore for consistency we have $\nabla_{-\omega} KN=0$.} and $\varphi$ a single--valued differential form. 
The twisted homology group $H_1(X,\Lc_\om)$ on $X$ comprises a space of twisted cycles  
$\Delta\otimes KN$ representing certain regions of $X$ subject to some additional information about branches of $KN$. Likewise, the twisted cohomology group $H^1(X,\nabla_\om)$ is the space of twisted differential forms (twisted cocycles), which are closed but not exact w.r.t.~$\nabla_\om$. Originally, intersection forms on twisted (co)homology groups have been developed for deriving hypergeometric  function identities involving twisted intersection numbers and eventually providing a unified description and generalization of hypergeometric integrals \cite{Gelfand:1990bua}.
Generalized hypergeometric integrals show up in tree--level  string amplitudes as integrals over world--sheet positions of vertex operators, cf.~\cite{Oprisa:2005wu}. 
Generically in this case we are dealing with multi--valued functions defined on the   moduli space $\mathcal{M}_{0,n}=\{(t_1,\ldots,t_i,\ldots,t_n) \in (\IC\IP)^{n-3}\ ,\ i\neq j,k,l \ |\  \mathop{\mathlarger{\forall}}_{m\neq n } t_m\neq t_n \}$ of Riemann spheres with $n$ punctures and an
$(n-3)$--simplex $\Delta$ referring to an ordering of the $n-3$ points $t_i,i\neq j,k,l$. Similarly, 
multi--valued functions on a complex torus with $n$ points describe 
string one--loop amplitudes. In particular, one basic example is given by the Riemann--Wirtinger integral 
defined on the moduli space ${\cal M}_{1,n}$ of $n$--punctured  elliptic curve $E_\tau\slash\{t_1,\ldots,t_n\}$ with $n$ distinct points $t_i$ on $E_\tau$.  At any rate, both at tree-- and one--loop level open string amplitudes are formulated as pairings \req{PAIR} between twisted cycles and cocycles. On the other hand, closed string amplitudes  are described by pairings between two twisted  cocyles.

The intersection form  $\vev{\ldots|\ldots}$ of twisted cycles on a one--dimensional complex manifold $X$ is a bilinear form between the twisted homology $H_1(X,\Lc^\vee_\om)$ and the associated (locally finite) twisted homology $H_1^{lf}(X,\Lc_\om)$
with the local system $\Lc_\om$ and its dual $\Lc^\vee_\om$:
\be\label{Rogers}
\vev{\ldots|\ldots}:\ H_1(X,\Lc^\vee_\om) \times H_1^{lf}(X,\Lc_\om)\lra \IC\ .
\ee
The two spaces $H_1(X,\Lc^\vee_\om)$ and  $H_1^{lf}(X,\Lc_\om)$ are dual to each other. Besides, there is an isomorphism $H_1^{lf}(X,\Lc_\om)\simeq H_1(X,\Lc_\om)$ called the regularization ${\rm reg}_\om$. 
The intersection number $\vev{\sigma|\tau}$ between two cycles $\si\in H_1(X,\Lc^\vee_\om)$ and $\tau\in H_1^{lf}(X,\Lc_\om)$ is defined as integral over the corresponding
Poincar\'e dual cohomology classes $\theta_c([\si])\in H^1_c(X,\nabla_{-\om})$ and $\theta([\tau])\in H^1(X,\nabla_{+\om})$ as \cite{Kita}:
\be\label{INTERSECT}
\vev{\si|\tau}=\int_X\theta_c([\si])\wedge\theta([\tau])\ .
\ee
The intersection number \req{INTERSECT} does not receive contributions from the bulk of $X$ and can be computed by summing up the local intersection numbers $I_\nu$ at the intersecting points $\nu$. For expansions
\bea
\sigma&=&\ds\sum_i c_i\;\Delta_i\otimes KN\ ,\\
\tau&=&\ds\sum_j c_j\;\Box_j\otimes KN^{-1}\ ,
\eea
we obtain \cite{Kita}:
\be\label{INTERSECTION}
\vev{\si|\tau}=\sum_{\Delta_i\cap\Box_i=\{\nu_{ij}\}}=c_i\; c_j\; KN(\nu_{ij})\;KN^{-1}(\nu_{ij})\;I_\nu(\Delta_i,\Box_j)\ .
\ee
Thus, intersection numbers \req{INTERSECTION} are computed by combinatorial rules taking into account how the cycles intersect in the moduli space under consideration. In particular, the formula \req{INTERSECTION} can  be applied for both genus zero and one case \cite{goto2023intersection}. Furthermore, we should note that in principle twisted intersection numbers may be defined on $n$--dimensional complex manifolds $X$ between elements of $H_n(X,\Lc^\vee_\om)$ and $H_n^{lf}(X,\Lc_\om)$. Calculation of the multi--dimensional case is discussed later in this work.

Intersection numbers can be used for analytic continuation and basis expansions of integrals. Moreover,  they appear in expressing closed string amplitudes in terms of a pair of open string amplitudes. 
This requires a suitable holomorphic splitting of left-- and right--movers in order to establish an orthonormal  basis expansion of twisted cycles in both sectors. Hence, in the following we shall be concerned with 
homological splittings on the sphere and torus closed string world--sheets.

\sect{Homological splitting on the sphere}

On the sphere a splitting of complex integration into holomorphic and anti--holomorphic sectors is achieved 
by performing an analytic continuation of coordinates. This procedure has been proposed by Kawai, Lewellen and Tye \cite{klt} and is related to twisted Riemann's
period relations.

\subsection{Tree--level KLT formula and twisted intersection numbers}

The closed string amplitude on the sphere reads
\be\label{String0}
\Mc^{closed}_{n;0}=V_{CKG}^{-1}\lf(\prod_{r=1}^n\int_{\IC} d^2z_r\ri)\;\prod_{i<j}|z_i-z_j|^{\ap q_iq_j}\; (z_i-z_j)^{n_{ij}}\; \ (\bar z_i-\bar z_j)^{\tilde n_{ij}}\ ,
\ee
with  the inverse string tension $\ap$ and some integers $n_{ij},\tilde n_{ij}\in\IZ$. This amplitude describes the scattering of $n$ closed string states of external momenta $q_i\in\IR$ subject to momentum conservation $\sum_{i=1}^nq_i=0$. The following discussion on monodromies and analytic continuation does not depend on the specific values of the integers $n_{ij}$. Therefore, in the following  we shall assume  $n_{ij},\tilde n_{ij}=0$. Furthermore, we restrict to  massless external states, i.e.  $q_i^2=0$. 
On the sphere a homological splitting 
\begin{align}
 z_r&=z^1_r+iz^2_r\simeq \xi_r\ ,\nonumber\\
\bar z_r&=z^1_r-iz^2_r\simeq \eta_r\ ,
\end{align} 
is realized by analytic continuation of $z^2_i$, which gives rise to the KLT relations \cite{klt}:
\begin{align}
\Mc^{closed}_{n;0}&=V_{\rm CKG}^{-1}\lf(\prod_{r=1}^{n}\int_{-\infty}^\infty d\xi_r \int_{-\infty}^\infty d\eta_r\ri)\!\!
\lf(\prod_{i<j}\Pi(\xi_i,\eta_i;\xi_j,\eta_j)\;(\xi_i-\xi_j)^{\h\ap q_iq_j}\;(\eta_i-\eta_j)^{\h\ap q_iq_j}\ri)\nonumber\\
&=\sum^\prime_{\al,\bet\in S_{n-1}} e^{i\Phi(\al,\bet)}\ A^{open}_{n;0}(\al)\; \tilde A^{open}_{n;0}(\bet)\ .\label{KLTtree0}
\end{align}
In Eq. \req{KLTtree0} the primed sum runs over all $\h(n-1)!$ cyclic invariant permutations $\al,\bet$ with $n-1$ to the right of $1$.
Furthermore, there is some phase factor $\Pi$ 
\be\label{Phases}
\Pi(r,s):=\Pi(\xi_s,\xi_r,\eta_s,\eta_r; q_rq_s)=e^{\h\pi i\ap q_rq_s (1-\theta[(\xi_r-\xi_s)(\eta_r-\eta_s)])}\ ,
\ee
rendering the correct branch cut structure and the open string subamplitudes
\bea\label{openAmp}
A^{open}_{n;0}(\al)&=&\ds V_{\rm CKG}^{-1}\int_{{\cal I}_\al}\lf(\prod_{s=1}^n dz_s\ri)\prod_{i<j} |z_i-z_j|^{\h\ap q_iq_j}\\
&\equiv& \ds|\Delta(\al)\otimes KN\rng=\int_{\Delta(\al)} KN\ ,
\eea
with the integration region ${\cal I}_\al$ (or cycle $\Delta(\al)$)
\be
{\cal I}_\al=\{ z_i \in\IR\;|\; z_{\al(1)}<\ldots<z_{\al(n)}\}\equiv \Delta(\al)\ ,
\ee
and similarly for $\tilde A^{open}_{n;0}$.  Eq. \req{KLTtree0} can be further reduced after selecting a basis of $(n-3)!$ independent cycles corresponding to  a set of subamplitudes $A^{open}_{n;0}(\al)$.

An alternative expression of \req{KLTtree0} can be given in terms of twisted intersection numbers~\cite{Mizera:2017cqs}
\begin{align}
\Mc^{closed}_{n;0}&=\sum^\prime_{\al,\bet\in S_{n-1}} \vev{\Delta(\al)\otimes KN|\Delta(\beta)^\vee\otimes 
\overline{KN}}\ A^{open}_{n;0}(\al)\; \tilde A^{open}_{n;0}(\bet)\ ,\label{KLTtree}
\end{align}
with the Koba--Nielsen factor and single--valued one--form:
\be\label{Pinnacles}
KN=\prod_{i<j}(z_i-z_j)^{\ap q_iq_j}\ \ \ ,\ \ \ \omega=d\ln KN\ .
\ee
Consequently, the intersection numbers \req{INTERSECTION} can be related to the monodromy phases appearing in 
\req{KLTtree0} as
\be\label{REL}
\vev{\Delta(\al)\otimes KN|\Delta(\beta)^\vee\otimes 
\overline{KN}}\simeq e^{i\Phi(\al,\bet)}\ ,
\ee 
subject to the identification $\overline{KN}\simeq KN^{-1}$ leading to the canonical isomorphism
$\Lc_{-\omega}\simeq \Lc_{\ov \omega}$ of local systems for real momenta $q_i$. 

\subsection{One complex sphere integration and $n$ unintegrated points}

\begin{figure}[H]
    \centering
    \includegraphics[scale=.45]{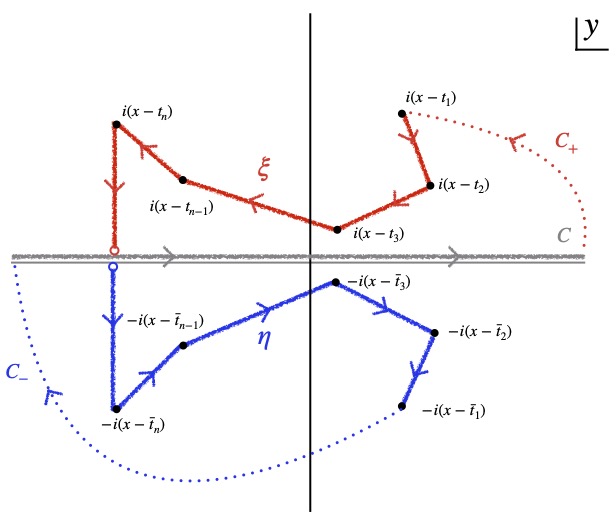}
    \caption{Branch points and closed contour in the complex $y$--plane.}
    \label{KLTtreeFig}
\end{figure}

Let us now discuss the case with one complex integration and $n$ unintegrated points $t_i\in\IC$:
\be\label{Start10}
\Mc^{closed}_{1;0}=V_{CKG}^{-1}\int_{\IC} d^2z\;\prod_{r=1}^n|z-t_r|^{2c_r}\ .
\ee
For the parameterization $z=x+iy$ (with $d^2z=\tfrac{i}{2}dx dy$) the integrand
\be\label{integrand}
I(x,y)=\prod_{r=1}^n (x+iy-t_r)^{c_r}\ (x-iy-\bar t_r)^{c_r}
\ee 
becomes an analytic function of $y$. The latter has $n$ pairs of branch points at $y=i(x-t_k)=i(x-\Re t_k)+\Im t_k$ and $y=-i(x-\bar t_k)=-i(x-\Re t_k)+\Im t_k$ with monodromy phases $e^{\pm\pi i c_k}$, where the factors $(z-t_k)^{c_k}$ and $(\bar z-\bar t_k)^{c_k}$ have zeros, respectively. For a given $x\in(-\infty,\infty)$ the location of these points is depicted in  Fig.~\ref{KLTtreeFig}.
To determine the integral \req{Start10} along $y\in(-\infty,\infty)$ we may perform an analytic continuation in $y$ by considering a contour in the complex $y$--plane. This contour passes all $2n$ branch points and is comprised by the three pieces  $C_+,C_-$ and $C$, depicted in Fig.~\ref{KLTtreeFig}.
\noindent 
 While $C_+$ connects all $n$ holomorphic branch points in the upper half--plane, the path $C_-$ passes all $n$ anti--holomorphic points in the lower half--plane. 
By Cauchy's theorem we can write for a given $x\in C_0$ with $C_0=\{x\;|\;x\in\IR\}$:
\be\label{cauchy}
\int_C dy\ I(x,y)+\int_{C_+}dy\ \hat I(x,y)+\int_{C_-}dy\ \hat I(x,y)=0\ .
\ee
Above, the object $\hat I$ refers to the expression \req{integrand}  rendered to be single--valued
when moving away from $C$. The $y$--integral of  \req{Start10} is described by the edge $C$ and with \req{cauchy} we may express it as:
\begin{align}
\int_C dy\ I(x,y)&=\int_{-\infty}^{\infty}I(x,y)\nonumber\\
&=-\int_{C_+}dy\ \hat I(x,y)-\int_{C_-}dy\ \hat I(x,y)=-\int_{C_+\cup C_-}dy\ \hat I(x,y)\ .\label{cauchy1}
\end{align}
With \req{cauchy1} the $y$--integral of  \req{Start10} can  be  expressed through the combination of cycles $C_+\cup C_-$.
Eventually, the two cycles $C_+$ and $C_-$ can be separated by pulling their finite ends to infinity or likewise joined to a single cycle $C_+\cup C_-$. The two cycles $C_+$ and $C_-$ are complex conjugate\footnote{For real parameterization of a cycle $\gamma$ with $\bar\gamma(t)=\overline{\gamma(t)}, t\in (a,b)$ we have: $\lf(\int_\gamma f(z)\;dz\ri)^\ast =\int_{\bar\gamma}\overline{f(\bar z)} \; dz$.} subject to orientation reversal, i.e. $C_-\simeq
-(C_+)^\ast$. Therefore, we also have~\cite{Stieberger:2022lss}
\be\label{hilfe}
\int_{C_-}dy\ \hat I(x,y)=-\lf(\int_{C_+}dy\ \hat I(x,y)\ri)^\ast\ ,
\ee
which guarantees the integral \req{cauchy1} to be purely  imaginary. Note, that in the next step, after defining proper real coordinates, the complete integral \req{Start10} becomes real.

For a given $x\in(-\infty,\infty)$ we may introduce the pair of complex coordinates
\be\label{changecoord}
\ba{lcl}
\xi&=&x-\tilde y\ ,\\
\eta&=&x+\tilde y\ ,
\ea\ee
with $\tilde y=-iy$.
These coordinates \req{changecoord} become real  along the imaginary axis $y\in i\IR$, i.e.~$\xi,\eta\in(-\infty,\infty)$. We may deform the latter to reach the paths $C_+$ and $C_-$ with  the $2n$ branch points. Combining according to \req{changecoord} the cycles $C_+$ and $C_-$  with the cycle $C_0$ for the $x$--integration we obtain two independent  integration paths $C_\xi$ and $C_\eta$.
Then, the contour $C_\xi=C_+\cup C_0$ (depicted in red) probes the $n$ zeros of $(\xi-t_i)$, while the contour $C_\eta=C_-\cup C_0$ (depicted in blue) hits the $n$ zeros of $(\eta-\bar t_i)$. 
As a consequence  we can write \req{Start10} in terms of the new coordinates \req{changecoord} and their corresponding cycles
\be\label{ContourKLT}
\Mc^{closed}_{1;0}=-V_{CKG}^{-1}\oint_{C_\xi}d\xi \prod_{r=1}^n(\xi-t_r)^{c_r}\ \oint_{C_\eta}d\eta \prod_{s=1}^n(\eta-\bar t_s)^{c_s}\ \Pi(\xi,\eta)\ ,
\ee
with the phases $\Pi(\xi,\eta)$ introduced in \req{Phases} rendering the integrand single--valued.
In other words, the phase factors $\Pi$ make sure, that we stay in the correct branch when the coordinates $\xi,\eta$ are varied along $C_+$ and $C_-$ between different segments $(t_i,t_{i+1})$ and  $(\ov t_k,\ov t_{k+1})$, respectively.

We can divide the integrations $\xi,\eta$ into a double sum over $n$ cycles $\gamma_{ij}$, which respect the orderings of the  unintegrated $n$ points $t_r$. With $\gamma_{ij}$ representing the twisted cycle  $\gamma_{ij}\simeq{\rm reg.}(t_i,t_j)$ from $t_i$ until $t_j$, with $j=i+1$ and $i=1,\ldots,n$ we have:
\be\label{ContourKLTa}
\Mc^{closed}_{1;0}=-V_{CKG}^{-1}\sum_{i,k=1}^n \int_{\gamma_{i,i+1}}dz\prod_{r=1}^n(z-t_r)^{c_r}\ \int_{\gamma_{k,k+1}}d\bar z \prod_{s=1}^n(\bar z-\bar t_s)^{c_s}\ \Pi(i,k)\ .
\ee
Then, in \req{ContourKLTa} the phase $\Pi(i,k)$  has the interpretation of twisted intersection number \be
\Pi(i,k)=\vev{\gamma_{i,i+1}|\gamma_{k,k+1}}\ ,
\ee
in lines of \req{REL}.

\subsection{Multi--dimensional integration on the sphere}

Let us now discuss the case of multi--dimensional complex integration on the sphere leading to \req{KLTtree0} and \req{KLTtree}. We want to start from the results \req{ContourKLT} or \req{ContourKLTa} supplemented by the additional factors
$$
\prod_{i<j}^n(t_i-t_j)^{c_{ij}}\; (\bar t_i-\bar t_j)^{c_{ij}}\ ,
$$
with $c_{ij}=c_{ji}$ and systematically perform complex integrations w.r.t.~all $n$ coordinates $t_r$. Note, that in the end this computes $\Mc^{closed}_{n+1;0}$.
We start at $t_n$ and consider the relevant integrand
$$
(\xi-t_n)^{c_n}\;(\eta-\bar t_n)^{c_n}\;\prod_{i=1}^{n-1}(t_i-t_n)^{c_{in}}\; (\bar t_i-\bar t_n)^{c_{in}}
$$
as analytic function in $\Im t_n$, with $t_n=\Re t_n+i\Im t_n$.
In the complex $\Im t_n$--plane there are $n$ pairs of branch points.
One pair at $\Im t_n=-i(\xi-\Re t_n)$ and $\Im t_n=i(\eta-\Re t_n)$ with phase $e^{\pi i c_{n}}$ and $n-1$ pairs at $\Im t_n=-i(t_l-\Re t_n)$ and $\Im t_n=i(\bar t_l-\Re t_n),\ l=1,\ldots,n-1$ with corresponding phases $e^{\pi i c_{ln}}$. In the complex $\Im t_n$--plane the locations of these branch points is similar  to the case studied before and depicted in Fig.~\ref{KLTtreeFig}. Only the pair of points $\Im t_n=-i(\xi-\Re t_n)$ and $\Im t_n=i(\eta-\Re t_n)$ is not complex conjugate to each other and the latter are already aligned along the imaginary axis of $\Im t_n$.
Nevertheless, we may apply the previous steps and consider two cycles $C_\pm$ with $C_+$ connecting all the $n$ branch points $\Im t_n=-i(\xi-\Re t_n)$ and $\Im t_n=-i(t_l-\Re t_n)$, while $C_-$ passing through the $n$ points $\Im t_n=i(\eta-\Re t_n)$ and $\Im t_n=i(\bar t_l-\Re t_n)$.
For a given $\Re t_n\in(-\infty, \infty)$ we introduce the two independent new complex coordinates 
\be\label{changecoordn}
\ba{lcl}
\xi_n&=&\Re t_n+i\Im t_n\ ,\\
\eta_n&=&\Re t_n-i\Im t_n\ ,
\ea\ee
which become real along the imaginary axis of $\Im t_n$, i.e. for $\Im t_n\in i\IR$.
Again, according to \req{changecoordn} the cycles $C_+$ and $C_-$ are combined with 
the cycle describing the $\Re t_n$--integration giving rise to two independent integration paths 
$C_{\xi_n}$ and $C_{\eta_n}$, with $C_{\xi_n}$ probing the $n$ zeros of $(\xi-\xi_n)$ and $(t_l-\xi_n)$ and $C_{\eta_n}$ hitting the zeros of $(\eta-\eta_n)$ and $(\bar t_l-\eta_n)$, respectively.
In addition to the $n$ cycles $\gamma_{i,i+1}$ for the $z$-integration we now have the $n-1$ cycles 
describing the $t_n$ integration between the $n-1$ points $t_1,\ldots,t_{n-1}$ thus in total there are $n(n-1)$ terms.
Eventually, after successively applying the above steps to all (except three) coordinates $t_l$ we arrive at \req{KLTtree} with $(\h n!)^2$ terms. 

Twisted cycles and cocycles for configuration space  integrals over punctured Riemann spheres (with $p$ integrated points and $n-p$ punctures) have also been discussed in \cite{Britto:2021prf}.
Let us conclude by pointing out
that the holomorphic splitting  \req{KLTtree0} of the complex sphere integrations \req{String0} may be stated by   inserting \cite{MT}
\be\label{Saguaro}
{\bf 1}=\sum^\prime_{\si,\rho\in S_{n-1}}e^{i\Phi(\si,\rho)}\;|\Delta(\si)\otimes \overline{KN}\rng\ \lng\Delta(\rho)\otimes KN|\ 
\ee
to arrive at \req{KLTtree}  with the local system \req{Pinnacles} and the subamplitudes \req{openAmp}.

\sect{Riemann--Wirtinger integral and twisted intersection numbers}
\label{RWIsection}

The Riemann--Wirtinger integral is introduced  as an analogue  of the hypergeometric integral representation. It is defined  as  an integral over some (twisted) cycle $\gamma$ on a one--dimensional complex torus  
torus $E_\tau={\bf C}/\Lambda$ with the lattice $\Lambda$ generated by $1$ and $\tau$ \cite{Mano,Mano2}, cf. also \cite{Watanabe,goto2023intersection,ghazouani2016moduli}:
\be\label{RiWi}
\int_\gamma\; T(z)\; dz\ .
\ee 
Above for a given $z\in\IC$ we have the local system with the multi--valued holomorphic function
\be\label{Wendelstein}
T(z)= e^{2\pi i c_0 z}\prod_{k=1}^n \theta_1(z-z_k;\tau)^{c_k}\equiv KN_z\ ,
\ee
with $c_0\in\IC$ and $c_i\in \IC\backslash\IZ$ subject to:
\be
\sum_{i=1}^n c_i=0\ .
\ee
The parameter $c_\infty$ arises from
\be\label{Periodtau}
T(z+\tau)=e^{2\pi i c_\infty}\;T(z)\ ,
\ee
\be\label{cinf} 
c_\infty=c_0\;\tau+\sum_{i=1}^nc_iz_i\ .
\ee
Furthermore, we have:
\be\label{Phase1}
T(z\pm1)=e^{\pm2\pi i c_0}\;T(z)\ .
\ee
Later we will set
\be\label{CHOICE}
c_0=-\h \ap\ell q_{n+1}\ \ \ ,\ \ \ c_k=\h \ap q_{n+1}q_k\ 
\ee
to describe the $z$--integrand of an  $n+1$--point amplitude, cf. Subsection \ref{KLTDez}.

The hypergeometric integral representation can be translated into a pairing of twisted homology class and twisted cohomology class on the complex sphere  minus some points. 
In this Section we study the intersection forms on the twisted homology and cohomology groups associated with the Riemann--Wirtinger integral \req{RiWi}.
There are the local system and dual local system 
\begin{align}
\Lc_\om:&=\Lc_\om(c_0,c_1,\ldots,c_n)=\IC\; KN_z\ ,\label{L1}\\
\Lc^\vee_{-\om}:&=\Lc^\vee_{-\om}(-c_0,-c_1,\ldots,-c_n)=\IC\; KN_z^{-1}\ ,\label{L2}
\end{align}
associated to the multi--valuedness  of the twist
\be
\om=d\ln KN_z
\ee
and dual twist, respectively. These local systems are related by the involution $c_i\ra -c_i$, which
lead to the construction of the twisted cohomology group $H^1(X,\nabla_\om)$ and homology group $H_1(X,\Lc_{\om})$ of the Riemann--Wirtinger integral and likewise their duals $H^1(X,\nabla_{-\om})$ and  $H_1(M,\Lc^\vee_{\om})$, respectively. Furthermore, \req{L1} and \req{L2}  are line bundles, which capture the monodromy properties of the twist and dual twist, respectively.

In the following we shall assume the gauge choice $z_1\!=\!0$. On the torus there are $n\!+\!1$ cycles $\gamma_{1\infty},\gamma_{10},\gamma_{12},\ldots,\gamma_{1n}$ with the natural map (Veech's full holonomy map) or pairing
\be\label{OPEN}
a^{open}_{n;1}(\gamma_i)=\int_{\gamma_i} dz\;T(z)
\ee
giving rise to elliptic hypergeometric integrals. Note, that there exist other versions of Rie\-/mann--Wirtinger integral representations involving 
the quasi--periodic Kronecker--Eisenstein series and \req{OPEN} can be obtained as natural limit thereof \cite{ghazouani2016moduli}.
Again, the set of $n+1$ cycles $\gamma$ satisfies a single $\IC$--linear monodromy relation \cite{Mano}:
\be\label{Monodromy}
\sum_{j=2}^ne^{-2\pi i (c_1+\ldots+c_j)}\;(1-e^{2\pi i c_j})\;\gamma_{1j}+(1-e^{2\pi i c_0})\;\gamma_{1\infty}=(1-e^{-2\pi i c_\infty})\;\gamma_{10}\ .
\ee
We can choose a basis of $H_1$ comprising $n$ cycles $\gamma_j\equiv \gamma_{1j}={\rm reg}(t_1,t_j)\; j=2,\ldots,n$ and $\gamma_0\equiv\gamma_{10}={\rm reg}(t_1,t_1+1)$, with the identification $z_k\!=\!t_k$. Then the relation  \req{Monodromy} allows to eliminate e.g.~the cycle $\gamma_{1\infty}$ by expressing it in terms of the remaining $n$ cycles $\gamma_{10},\gamma_{12},\ldots,\gamma_{1n}$. Note, that for convenience we may use 
$\gamma_{jk}={\rm reg}(t_j,t_k)=\gamma_{1k}-\gamma_{1j}\equiv\gamma_k-\gamma_j$. Furthermore, with \req{OPEN} the relation \req{Monodromy} gives rise to an identity between periods. From the physical point of view the 
monodromy relation \req{Monodromy} appears as world--sheet string monodromy 
relating  open string one--loop cylinder subamplitudes \cite{Hohenegger:2017kqy,Tourkine:2016bak}.
Likewise, the corresponding monodromy relation on the  doubled surface has been established in \cite{Stieberger:2022lss}.

We consider the following chain of $n$ twisted cycles
\be\label{Cycles}
\gamma_{12},\gamma_{23}\ ,\ldots,\gamma_{n-1,n},\gamma_{n,0}\ ,
\ee
with 
\be\label{defn,0}
\gamma_{n,0}=\gamma_n-\gamma_0\equiv\gamma_{10}-\gamma_{1n}
\ee
depicted in the next figure
\begin{figure}[H]
    \centering
    \includegraphics[scale=.55]{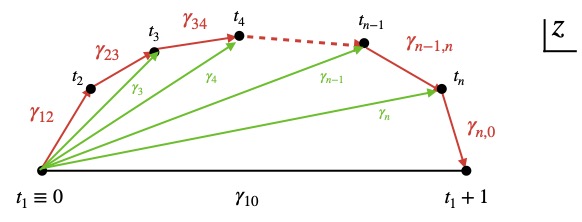}
    \caption{Chain of $n$ cycles}
    \label{Chain of cycles}
\end{figure}
\noindent  and subject to the following monodromy relation:
\be\label{Monodromy1}
\gamma_n-\sum_{i=2}^ne^{-2\pi i \sum\limits_{j=1}^{i-1}c_{j}}\;\gamma_{i-1,i}+(1-e^{2\pi i c_0})\;\gamma_{1\infty}=(1-e^{-2\pi i c_\infty})\;\gamma_{10}\ .
\ee
With 
\be\label{Richtig}
\gamma_n\equiv\gamma_{1n}=\sum\limits_{i=2}^n\gamma_{i-1,i}
\ee
the relation \req{Monodromy1} allows to express $\gamma_{1\infty}$ in terms of the $n$ cycles $\gamma_{10},\gamma_{12},\ldots,\gamma_{n-1,n}$ or likewise with \req{defn,0} in terms of the $n$ cycles $\gamma_{12},\ldots,\gamma_{n-1,n},\gamma_{n,0}$.
Hence, \req{Cycles} represents a basis of cycles. The intersection properties of the $n-1$ cycles $\gamma_{12},\gamma_{23}\ ,\ldots,\gamma_{n-1,n}$ 
coincides with those \req{KLTtree} stemming from the function $(z-x_1)^{c_1}(z-x_2)^{c_2}\cdot\ldots\cdot
(z-x_n)^{c_n}$ on $\IP^1-\{x_1=\infty,x_2,\ldots,x_n\}$ \cite{Kita}. 
Furthermore, we have the self--intersection number
\be\label{Int12}
I_h([\gamma_{12}],[\gamma_{12}^\vee])=\fc{d_{12}}{d_1d_2}\ ,
\ee
with $d_{ij}=e^{2\pi i(c_i+c_j)}-1$.
In addition, we can verify and compute:
\be\label{Intn0}
I_h([\gamma_{n,0}],[\gamma_{n,0}^\vee])=I_h([\gamma_{1n}],[\gamma_{1n}^\vee])=
\fc{1-e^{2\pi i (c_1+c_n)}}{(1-e^{2\pi i c_1})(1-e^{2\pi i c_n})}\equiv \fc{d_{1n}}{d_1d_n}\ .
\ee
Note, that the self--intersection numbers \req{Int12}  and \req{Intn0}  are the same as on the sphere. The same is true
for the following intersection numbers:
\begin{align}
I_h([\gamma_{n,0}],[\gamma_{n-1,n}^\vee])&=I_h([\gamma_{10}],[\gamma_{n-1,n}^\vee])-I_h([\gamma_{1n}],[\gamma_{n-1,n}^\vee])=
                                          \fc{e^{2\pi i c_n}}{d_n}\ ,\label{Int10n-1}\\
I_h([\gamma_{n-1,n}],[\gamma_{n,0}^\vee])&=I_h([\gamma_{n-1,n}],[\gamma_{10}^\vee])-I_h([\gamma_{n-1,n}],[\gamma_{1n}^\vee])
                                         = \fc{1}{d_n}\ ,\label{Int0n-11}\\
    I_h([\gamma_{n,0}],[\gamma_{jk}^\vee])&=0\ ,\  jk\neq12\ \mbox{and:}\ jk\neq n-1,n \ .                                 
\end{align}
On the other hand, since we have
\be
I_h([\gamma_{12}],[\gamma_{10}^\vee])=\fc{e^{2\pi i c_1}(1-e^{-2\pi i c_0})}{1-e^{2\pi i c_1}}\ ,
\ee
and $\gamma_{n,0}$ is composed of $\gamma_{10}$  we have the non--standard intersection number
with an additional second term, which is absent on the sphere:
\begin{align}
I_h([\gamma_{n,0}],[\gamma_{12}^\vee])&=-I_h([\gamma_{1n}],[\gamma_{12}^\vee])+I_h([\gamma_{10}],[\gamma_{12}^\vee])\nonumber\\
&=-\fc{1}{1-e^{2\pi i c_1}}+\fc{1-e^{2\pi i c_0}}{1-e^{2\pi i c_1}}=-
e^{2\pi i c_0}\   \fc{1}{1-e^{2\pi i c_1}}=
e^{2\pi i c_0}\ \fc{1}{d_1}\ ,\label{Intn012}\\
  I_h([\gamma_{12}],[\gamma_{n,0}^\vee])&=-I_h([\gamma_{12}],[\gamma_{1n}^\vee])+I_h([\gamma_{12}],[\gamma_{10}^\vee])\nonumber\\
&=-\fc{e^{2\pi i c_1}}{1-e^{2\pi i c_1}}+\fc{e^{2\pi i c_1}(1-e^{-2\pi i c_0})}{1-e^{2\pi i c_1}}=-e^{-2\pi i c_0}\ 
\fc{e^{2\pi i c_1}}{1-e^{2\pi i c_1}}=e^{-2\pi i c_0}\ 
\fc{e^{2\pi i c_1}}{d_1}\ .\label{Int12n0}
\end{align}
Hence,  the intersection numbers \req{Intn012} and \req{Int12n0} develop the additional factors $e^{2\pi i c_0}$ and $e^{-2\pi i c_0}$, respectively in contrast to the sphere case. In other words,  passing the cycle $\gamma_{10}$ from 
left amounts to the additional phase factor $e^{-2\pi i c_0}$. The difference to the sphere case
may also be understood  from the fact that in this case we may choose $t_1\simeq\infty$ such that $\gamma_{10}$ is effectively  pulled to infinity and hence not contributing.
Note, that we may redefine $\gamma_{n,0}$ (and $\gamma_{n,0}^\vee$) by $\tilde\gamma_{n,0}:=e^{2\pi i c_0}\gamma_{n,0}$ (and $\tilde\gamma_{n0}^\vee:=e^{-2\pi i c_0}\gamma_{n0}^\vee$) to cast \req{Intn012} and \req{Int12n0} into
the  form on the sphere albeit this changes \req{Int10n-1} and \req{Int0n-11}. Finally, we have:
\be
I_h([\gamma_{n,0}],[\gamma_{10}^\vee])=\fc{1-e^{2\pi i c_0}}{1-e^{2\pi i c_1}}\ .
\ee
Actually, in \cite{Casali:2019ihm} (cf. Section 3.3) it has been argued that the corner contributions of the two cycles $\gamma_{10}$ and $\gamma_{n,0}$ should be paired up into to the single twisted cycle
\be\label{Mizera}
\gamma_{1n}=\gamma_{10}-\gamma_{n,0}\ ,
\ee
with standard intersection numbers with the cycles $\gamma_{12},\ldots,\gamma_{n-1,n}$.
Finally, with \req{Richtig} we also have the intersection number ($i=1,\ldots,n-1$):
\begin{align}
I_h([\gamma_{1n}],[\gamma^\vee_{i,i+1}])=\sum_{j=1}^{n-1}I_h([\gamma_{j,j+1}],[\gamma^\vee_{i,i+1}])
=\begin{cases}  -\tfrac{1}{d_1}, &i=1,\\
-\tfrac{c_n}{d_n}, & i=n-1,\\
0, &i \neq 1,n-1,
\end{cases}  \label{1n1}\\ 
I_h([\gamma_{i,i+1}],[\gamma_{1n}^\vee])=\sum_{j=1}^{n-1}I_h([\gamma_{i,i+1}],[\gamma^\vee_{j,j+1}])=\begin{cases}  -\tfrac{c_1}{d_1}, &i=1,\\
 -\tfrac{1}{d_n},& i=n-1,\\
0, &i \neq 1,n-1,
\end{cases}  \label{1n2}\\
I_h([\gamma_{1n}],[\gamma^\vee_{1n}])=\sum_{i,j=1}^{n-1}I_h([\gamma_{i,i+1}],[\gamma^\vee_{j,j+1}])= -\fc{d_{1n}}{d_1d_n}\ , \label{1n3}
\end{align}
which share the same pattern as on the complex sphere.
The results \req{1n1}--\req{1n3} can also be anticipated by using the direct definition  of $\gamma_{1n}$, cf.  \ref{AppA}.

\sect{Homological splitting on the torus}

On the torus a splitting of complex integration into  holomorphic and anti--holomorphic sectors can  also be achieved 
by performing an analytic continuation of coordinates. This procedure albeit restricted to the case
$\Re\tau=0$ has been proposed in \cite{Stieberger:2022lss}.

\subsection{One complex torus integration and $n$ unintegrated points}\label{OneP}

We are interested in the complex integral
\be\label{Startn1}
M=\int_T d^2 z\ T(z)\ \overline{T(z)}\ ,
\ee
where the function $T$ has been defined in   \req{Wendelstein}. In order for the integrand to be invariant  under the $B$--cycle shift $z\ra\tau$ according to \req{Periodtau}
we obtain the following constraint for \req{cinf}~\cite{ghazouani2016moduli}:
\be\label{CONSTR}
\Im c_\infty=0\ \ \ \Longrightarrow\ \ \ c_0=-\fc{\sum\limits_{i=1}^n c_i\Im z_i}{\tau_2}\ .
\ee
On the other hand, relaxing the constraint \req{CONSTR} can be achieved by introducing the loop momentum \req{CHOICE}. This is what in the following we shall be interested in, i.e. assuming $c_0$ not to depend on the positions $z_k$ as in \req{CONSTR}, but as in \req{CHOICE} on a loop momentum $\ell$ to be integrated over, cf. also Subsection \ref{KLTDez} for further explicit details.

Note, that the two cycles $\gamma_{1\infty},\gamma_{10}$ can be identified with a symplectic basis of the homology group of the torus with  $\gamma_{1\infty}$ corresponding to the $B$--cycle, while $\gamma_{10}$ to the $A$--cycle of the torus, respectively. 
Therefore, after introducing the torus coordinates (with $d^2z=\tau_2\ dx dy$ and $\tau_2=\Im\tau$) as
\be
z=x+\tau\; y\ \ \ ,\ \ \ x,y\in (0,1)\ ,
\ee
we have:
\be\label{Start}
M=\tau_2\int_0^1dx \int_0^1dy\ T(x+\tau y)\ \overline{T(x+\tau y)}\ .
\ee
Ultimately, we aim to convert the $B$--cycle integration into $A$--cycle integrations by 
means of analytic continuation.
For a homological splitting  
 we can  perform an expansion in the holomorphic and anti--holomorphic sector leading to
\be\label{Splitting}
M=\sum_{}\int_{\gamma_{ij}} d\xi\int_{\gamma_{kl}} d\eta\ \Omega(\xi,\eta)\ T(\xi)\ \ov T(\eta)\ ,
\ee
with some cycles to be specified below and a phase factor $\Omega$.

\begin{figure}[H]
    \centering
    \includegraphics[scale=.5]{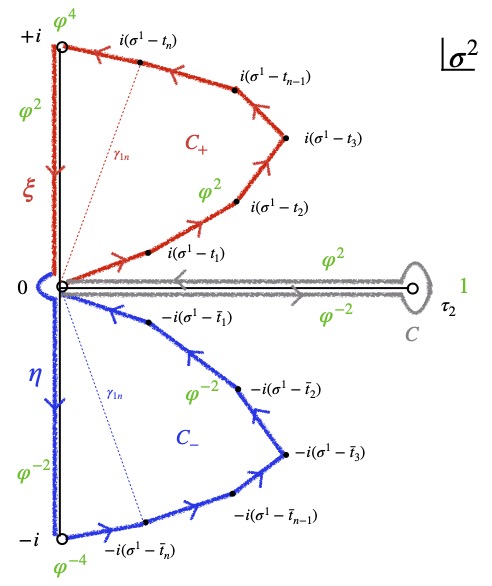}
    \caption{Complex $\sigma^2$--plane and closed contour for fixed $\sigma^1\in(0,1)$.}
    \label{Complex_Torus}
\end{figure}

Let us now specify the cycles contributing to the formula \req{Splitting}. According to \cite{Stieberger:2022lss} on the torus for\footnote{Actually, the condition $\Re\tau=0$ turns the equation $\overline{\theta_1(z,\tau)}=-\theta_1(\bar z,-\bar\tau)$ into $\overline{\theta_1(z,\tau)}=-\theta_1(\bar z,\tau)$. As a consequence for a given $\tau$ the local monodromy behaviour of the holomorphic 
sector $\prod_{k=1}^n \theta_1(z-z_k;\tau)^{c_k}$ is paired with that of the anti--holomorphic sector $\prod_{k=1}^n \theta_1(\bar z-\bar z_k;\tau)^{c_k}$ just like on the sphere.} $\Re\tau=0$  we can specify contours and apply Cauchy's theorem.
For the parameterization
\be
z=\si^1+i\si^2\ \ \ ,\ \ \ \si^1\in(0,1)\ ,\ \si^2\in(0,\tau_2)\ ,
\ee
with  $d^2z=d\sigma^1d\sigma^2$,  the integral \req{Start} becomes:
\be\label{SplittingA}
M=\int_0^1d\sigma^1\int_0^{\tau_2} d\sigma^2\ T(\sigma^1+i\sigma^2)\ \ov T(\sigma^1-i\sigma^2)\ .
\ee
To determine the integral \req{SplittingA} along $\sigma^2\in(0,\tau_2)$ we shall consider the single--valued integrand of \req{SplittingA} 
\be\label{Integrand}
I(\si^1,\si^2):=T(\si^1+i\si^2)\ \bar T(\si^1-i\si^2)
\ee
as holomorphic function in the complex $\sigma^2$--plane. This function has $n$ pairs of branch points  at $\sigma^2=i(\sigma^1-t_k)=i(\sigma^1-\Re t_k)+\Im t_k$ and $\sigma^2=-i(\sigma^1-i\bar t_k)=-i(\sigma^1-\Re t_k)+\Im t_k$ with monodromy phases $e^{\pm\pi i c_k}$,  where the factors $\theta_1(z-t_k)^{c_k}$ and $\theta_1(\bar z-\bar t_k)^{c_k}$ have zeros, respectively. For a given $\sigma^1\in(0,1)$ the location of these points is depicted in Fig.~\ref{Complex_Torus}. Incidentally, we can write the $c_0$ dependent factor of \req{Startn1} as
\be\label{Ventura}
e^{2\pi i c_0(z-\bar z)}=e^{-4\pi  c_0 \sigma^2}\sim\lf(\fc{\theta_1(i\sigma^2-\tau)}{\theta_1(i\sigma^2)}\ri)^{2c_0}
\ee
which behaves as $(\sigma^2)^{-2c_0}$ at $\sigma^2=0$ and as $(\sigma^2-\tau_2)^{2c_0}$ at $\sigma^2=-i\tau=\tau_2$. Hence, we find an additional pair of branch points with phases  $\varphi^{2}$ and $\varphi^{-2}$, respectively, with: 
\be
\varphi:=e^{-\pi i c_0}\ .
\ee
Likewise, we have
\be
e^{2\pi i c_0(z-\bar z)}= e^{-4\pi  c_0 \sigma^2}\sim\lf(\fc{\theta_1(i\sigma^2)}{\theta_4(i\sigma^2-\fc{\tau}{2})}\ri)^{-4c_0}=\lf(\fc{\theta_1(i\sigma^2+1)}{\theta_4(i\sigma^2+1-\fc{\tau}{2})}\ri)^{-4c_0}\ ,
\ee
exhibiting another  pair of branch points along $0\leq \Im(\sigma^2)\leq 1$ at $\sigma^2=0$ with the behaviour $(\sigma^2)^{-4c_0}$ and at $\sigma^2=i$ with the behaviour $(\sigma^2-i)^{-4c_0}$  each with phase $\varphi^{4}$.
The latter may also be anticipated by noting the shift symmetry:
\be\label{shiftS}
I(\si^1,\si^2\pm i)=\varphi^{\pm4}\ I(\si^1,\si^2\pm i)\ .
\ee
\noindent
To compute the integral \req{SplittingA} by following \cite{Stieberger:2022lss} we may perform an analytic continuation in $\sigma^2$ by considering a closed contour in the complex $\sigma^2$--plane
and apply Cauchy's theorem. This contour passes all $2n+4$ branch points (including $0,\tau_2,+i,-i$) and is composed by the three pieces  $C_+,C_-$ and $C$, depicted in Fig.~\ref{Complex_Torus}. The phase factor for the integrand of $C$ is now given by $\varphi^2$ and $\varphi^{-2}$ for the upper and lower contour respectively. While $C_+$ connects all $n$ holomorphic branch points in the upper half--plane, the path $C_-$ collects all $n$ anti--holomorphic points in the lower half--plane. 
By Cauchy theorem we can write for a given $\sigma^1\in(0,1)$:
\be\label{Cauchy}
\int_C d\sigma^2\ I(\si^1,\si^2)+\varphi^2\;\int_{C_+}d\sigma^2\ \hat I(\si^1,\si^2)+\varphi^{-2}\;\int_{C_-}d\sigma^2\ \hat I(\si^1,\si^2)=0\ .
\ee
Above, the object $\hat I$ refers to the expression \req{Integrand}  rendered to be single--valued
when moving away from $C$. The $\sigma^2$--integral of  \req{SplittingA} is described by the edge $C$ and with \req{Cauchy} we may express it as:
\begin{align}
\int_C d\sigma^2\ I(\si^1,\si^2)&=(-\varphi^2+\varphi^{-2})\int_0^{\tau_2}d\sigma^2\ I(\si^1,\si^2)\nonumber\\
&=-\varphi^2\int_{C_+}d\sigma^2\ \hat I(\si^1,\si^2)-\varphi^{-2}\int_{C_-}d\sigma^2\ \hat I(\si^1,\si^2)\ .\label{Cauchy1}
\end{align}
We have the relation between the integrals over the two cycles $C_+$ and $C_-$ \cite{Stieberger:2022lss}
\be\label{Hilfe}
\int_{C_-}d\sigma^2\ \hat I(\si^1,\si^2)=-\lf(\int_{C_+}d\sigma^2\ \hat I(\si^1,\si^2)\ri)^\ast\ ,
\ee
which is the one--loop analog of \req{hilfe}.
In addition, we can consider the two relations:
\begin{align}
(1-\varphi^2)\int_0^{\tau_2}d\sigma^2\ I(\si^1,\si^2)&+\varphi^2\; \int_{C_+}d\sigma^2\ \hat I(\si^1,\si^2)=0\ ,\label{sigl1}\\
(-1+\varphi^{-2})\int_0^{\tau_2}d\sigma^2\ I(\si^1,\si^2)&+\varphi^{-2}\; \int_{C_-}d\sigma^2\ \hat I(\si^1,\si^2)=0\ .\label{sigl2}
\end{align}
The two equations \req{sigl1} and \req{sigl2}, which thanks to \req{Hilfe} are complex conjugate to each other, can be combined to provide \req{Cauchy1}. Likewise, we find:
\be\label{SantaMonica}
\int_0^{\tau_2}d\sigma^2\ I(\si^1,\si^2)=(1-\varphi^{-2})^{-1}\int_{C_+}d\sigma^2\ \hat I(\si^1,\si^2)\ .
\ee
Note, that due to this relation considering only 
$C_+$ for the analytic continuation is enough, cf. also \cite{Stieberger:2022lss}.

For a given $\sigma^1\in C_0$, with $C_0=\{\sigma^1\;|\;\sigma^1\in(0,1)\}$ we may introduce the pair of complex coordinates
\be\label{Changecoord}
\ba{lcl}
\xi&=&\sigma^1-\tilde \sigma^2\ ,\\
\eta&=&\sigma^1+\tilde \sigma^2\ ,
\ea
\ee
with $\tilde \sigma^2=-i\sigma^2$.
These coordinates \req{Changecoord} become real along the imaginary axis $\sigma^2\in i\IR$, i.e. $\xi,\eta\in\IR$. 
We  may deform the latter to reach the paths $C_+$ and $C_-$ with  the $2n$ branch points. 
Combining the cycles $C_+$ and $C_-$  with the cycle $C_0$ for the $\si^1$--integration according to \req{Changecoord}  we obtain two independent  integration paths $C_\xi$ and $C_\eta$.
Then, the contour $C_\xi=C_+\cup C_0$ (depicted in red) probes the $n$ zeros of $\theta_1(\xi-t_i)$, while the contour $C_\eta=C_-\cup C_0$ (depicted in blue) hits the $n$ zeros of $\theta_1(\eta-\bar t_i)$. 
Recall, that in the case of the complex plane the two contours $C_+$ and $C_-$ can be decoupled by pulling their finite ends to infinity. This is not the case here, where the two contours $C_\eta$ and $C_\xi$   are connected at $\si^2=0$, i.e. $\xi=\eta$ resulting in an additional coordinate dependent  (splitting) function $\Psi'(\xi,\eta)$.
This function originates from the change of coordinates \req{Changecoord} and will be specified below.
Eventually, with \req{Cauchy1} and \req{SantaMonica} we obtain
\begin{align}
M&=\lf(1-e^{2\pi i c_0}\ri)^{-1}\ \oint_{C_\xi}d\xi\; T(\xi)\ \oint_{C_\eta}d\eta \;\bar T(\eta) \;\Psi'(\xi,\eta)\;\Pi(\xi,\eta)\ ,\nonumber\\
&=\lf(1-e^{2\pi i c_0}\ri)^{-1}\ \oint_{C_\xi}d\xi \prod_{r=1}^n\theta_1(\xi-t_r;\tau)^{c_r}\ \label{ContourKLTone}\\
&\times\oint_{C_\eta}d\eta\; e^{-2\pi i c_0 (\eta-\xi)}\;\Psi'(\xi,\eta)\;\prod_{s=1}^n\theta_1(\eta-\bar t_s;-\bar\tau)^{c_s}\; \Pi(\xi,\eta)\ ,\nonumber
\end{align}
with the phase $\Pi(\xi,\eta)$ rendering the integrand single--valued.
The latter is identical to the  phase $\Pi(\xi,\eta)$ introduced in  \req{ContourKLT}, which in turn has an    interpretation of twisted intersection numbers on the torus.

In fact, the monodromies of $e^{2\pi i c_0 \xi}\prod_{r=1}^n\theta_1(\xi-t_r;\tau)^{c_r}$ are related to the local system $\Lc_\om(c_i)$ given in \req{L1}. Likewise, the monodromies of $e^{-2\pi i c_0 \eta}\prod_{s=1}^n\theta_1(\eta-\bar t_s;-\bar\tau)^{c_s}$  agree with those of $\overline{\Lc_{\om}(c_i)}$, which in turn can be related to \req{L2} for real coefficients $c_i$ by the isomorphism $\Lc_{-\om}\simeq \Lc_{\bar \om}$ through complex conjugation. 
The two cycles $C_\xi$ and $C_\eta$ can be decomposed w.r.t.~the basis of $n$ cycles \req{Cycles} together with $\gamma_{10}$. Furthermore, by applying
\req{Mizera} it is convenient to replace the cycles $\gamma_{10}$ and $\gamma_{n,0}$ by $\gamma_{1n}$  (dashed line in Fig.~\ref{Complex_Torus}) and stay for both $C_\xi$ and $C_\eta$ on the branches $\varphi^{\pm 2}$  without 
passing $\si^2=\pm i$, respectively. 
This way the cycles $C_\xi$ and $C_\eta$ attain  the topology of  closed cycles passing through all $2n$ points $t_i,\bar t_i$ subject to $t_1,\bar t_1=0$, respectively.  
Consequently, we may consider the set of $n$ relevant cycles
\be\label{CyclesN}
\Gamma:=\{\gamma_{12},\gamma_{23}\ ,\ldots,\gamma_{n-1,n},\gamma_{1n}\}
\ee
and their corresponding intersection numbers discussed in the previous Section \ref{RWIsection}. Accordingly, the set of cycles \req{CyclesN} shares the same intersection properties than on the complex plane ${\bf P}^1-\{t_1,\ldots,t_n\}$.
Then, in \req{ContourKLTone} the phase $\Pi(\xi,\eta)$  can be expressed in terms of these twisted intersection numbers.
Eventually, the expression \req{ContourKLTone} assumes the form of  \req{Splitting} subject to
\be\label{Fully}
\Omega(\xi,\eta)\simeq \Psi(\xi,\eta)\ \Pi(\xi,\eta)\ ,
\ee
with the usual KLT phase factor $\Pi(\xi,\eta)$ together with the splitting function $\Psi(\xi,\eta):=\lf(1-e^{2\pi i c_0}\ri)^{-1}\Psi'(\xi,\eta)$   stemming from the change of coordinates \req{Changecoord}. Note, that the $n$ points $t_i$ and $\bar t_i$ are ordered along the paths $C_\xi$ and $C_\eta$, respectively.
For both sectors $\xi\in C_\xi$ and $\eta\in C_\eta$ this ordering gives rise to the configuration of $n$ cycles $\gamma_{i,i+1},\; i=1,\ldots,n$ depicted in Fig.~\ref{Chain of cycles}.
Their intersection properties have been studied in Section \ref{RWIsection}. Consequently, we have the   relationship:
\be\label{INTTorus}
\Pi(\xi,\eta)\simeq \vev{\gamma_{ij}\otimes KN_z|\gamma_{kl}^\vee\otimes 
  \overline{KN}_{\bar z}}\ .
\ee
More concretely with this identification Eq.~\req{ContourKLTone} can be cast into
\bea\label{LAX}
M&=&\ds\sum_{\gamma\in\Gamma}\sum_{\tilde\gamma\in\Gamma}\vev{\gamma\otimes KN_z|\tilde\gamma^\vee\otimes 
  \overline{KN}_{\bar z}}\ \int_\gamma dz\;  \prod_{r=1}^n\theta_1(z-t_r;\tau)^{c_r}\\
&\times&\ds\int_{\tilde\gamma} d\bar z\; e^{-2\pi i c_0 (\bar z-z)}\;\Psi(z,\bar z)\;\prod_{s=1}^n\overline{\theta_1( z- t_s;\tau)}^{c_s}\ .
\eea
According to \req{INTERSECT} the intersection numbers $\Pi$ are computed as complex (bulk) integrals over  the elliptic curve $X$ and localize  near the boundary of the moduli space describing configurations in which two or more punctures coalesce \cite{Kita}. For this description we 
require the local systems \req{L1} and \req{L2} to be related as $\overline{KN}_{\bar z}\simeq KN^{-1}_z$, which is true for real coefficients $c_i$. Likewise, the intersection numbers supply local properties stemming from the branching $\Pi(r,s)$, which appear after transforming to the coordinates  \req{Changecoord}. On the other hand, the splitting function $\Psi(\xi,\eta)$ describes the effects  related to the cutting of the torus $X$ and thus are not captured by \req{INTTorus}.
Note, that in order to deal in \req{Startn1} with an integrand invariant  under $B$--cycle shifts
one may take the choice \req{CHOICE} (with $\sum_{k=1}^n q_{n+1}q_k=0$) and integrate $M$ over the loop momentum $\ell$, cf. also Subsection  \ref{KLTDez}.

If all remaining unintegrated points $t_i$ were located along the imaginary axis, i.e. $\Re t_i=0$ 
there is an explicit expression for the splitting function \cite{Stieberger:2022lss}:
\be\label{SPLIT}
\Psi(\xi,\eta)=\fc{\lf(1+e^{2\pi ic_0}\ri)}{\lf(1-e^{2\pi ic_0}\ri)}\; e^{-2\pi i c_0\theta(\eta-\xi)}\ .
\ee
For this case the contours $C_\xi$ and $C_\eta$ can be located along the unit segment $\xi,\eta\in(0,1)$ \cite{Stieberger:2022lss}.

\begin{figure}[H]
    \centering
    \includegraphics[scale=.45]{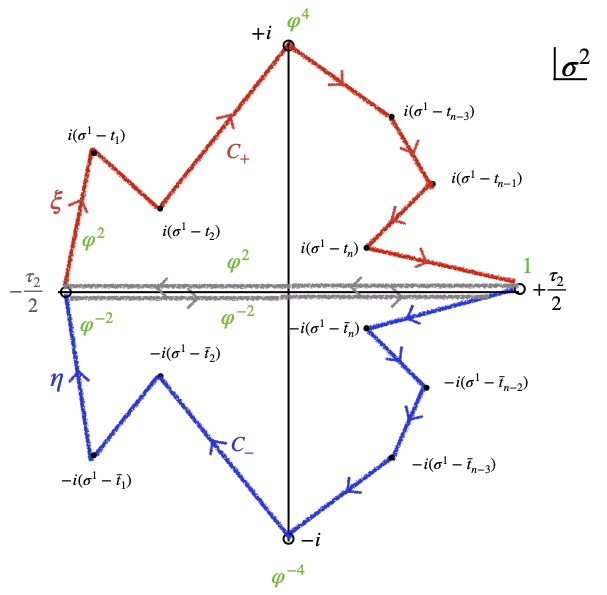}
    \caption{Complex $\sigma^2$--plane and alternative closed contour for fixed $\sigma^1\in(0,1)$.}
    \label{Complex_Torus1}
\end{figure}

Note, that instead  of the contours depicted in Fig.~\ref{Complex_Torus} we could also 
discuss the alternative closed contours shown in Fig.~\ref{Complex_Torus1}.
\noindent
These contours follow from the discussion in Appendix A of \cite{Stieberger:2022lss}.
The branch cut structure at $\si^2=\pm\tfrac{\tau_2}{2}$ with phases $\varphi^{\mp2}$  can be anticipated from
\be
e^{2\pi i c_0(z-\bar z)}=e^{-4\pi  c_0 \sigma^2}=\lf(\fc{\theta_1(i\sigma^2-\fc{\tau}{2})}{\theta_1(i\sigma^2+\fc{\tau}{2})}\ri)^{2c_0}\ ,
\ee
replacing \req{Ventura} and at $\si^2=\pm i$ from \req{shiftS}. The contour $C_+$ encircles the points $\si^2=\pm\tfrac{\tau_2}{2}$ by a semi--arc giving rise to the  phase factors $\varphi^{\pm2}$, respectively. On the other hand, the point $\si^2=i$ is passed by a quarter arc giving rise to the additional phase factors $\varphi^{\pm2}$, respectively.
In total we obtain:
\be\label{Nussbaum}
(-\varphi^2+\varphi^{-2})\int\limits_{-\tau_2/2}^{\tau_2/2}d\sigma^2\; I(\si^1,\si^2)
+\varphi^2\int_{C_+}d\sigma^2\; \hat I(\si^1,\si^2)+\varphi^{-2}\int_{C_-}d\sigma^2\; \hat I(\si^1,\si^2)=0\ .
\ee
However, to account for the correct monodromy phases in \req{Splitting} the cycle $\gamma_{1\infty}$ 
is also used.
An other note is, that instead performing an analytic continuation in the variable $\si^2$ we could also consider contours in the complex $\si^1$--plane. Then, the final result would 
assume a similar form than \req{ContourKLTone}, in particular with $c_0$ replaced by $c_\infty$.

\subsection{Multi--dimensional complex integration on the torus}\label{GENn}

Let us now discuss the case of multi--dimensional complex integration on the torus by starting at
\req{Startn1} and integrate w.r.t.~the remaining $n$ unintegrated points $t_i$. We shall consider
\bea\label{ConsiderN}
M&=&\ds V_{CKG}^{-1}\ \lf(\prod_{i=1}^n\int_T d^2 t_i\ri)\int_T d^2 z\ T(z)\ \overline{T(z)}\\[5mm]
&\times&\ds e^{2\pi i \sum\limits_{l=1}^n c_{0l} (t_l-\bar t_l)}\ \prod_{i<j}^n\theta_1(t_i-t_j)^{c_{ij}}\; \theta_1(\bar t_i-\bar t_j)^{c_{ij}}\ ,
\eea
with $c_{ij}=c_{ji}$ and $\sum_{i=1}^n c_{ij}=0$. This is the relevant case to write the one--loop KLT relation \cite{Stieberger:2022lss} as twisted period relation after allowing for a  loop momentum dependence of the parameters $c_{0l}$ and integrating over the loop momentum to guarantee invariance of the integrand under $B$--cycle shifts. In fact, only $n-1$ points need to be integrated  over the torus due to the fixing $t_1=0$. This procedure cancels the volume  $V_{CKG}=\tau_2$ of the conformal Killing group on the torus. 

The result of the $z$--integration can be found  in  \req{ContourKLTone} 
 supplemented by the additional $z$--independent factors.
Now we systematically perform complex integrations w.r.t.~to all $n$ coordinates $t_r$. We start at $t_n$ and consider the holomorphic functions
\bea\label{Siglalm}
T_n(t_n)&=&\ds e^{2\pi i c_{0n} t_n}\ \theta_1(\xi-t_n)^{c_n}\;\prod_{i<n}\theta_1(t_i-t_n)^{c_{in}}\ ,\\ 
\overline{T_n(t_n)}&=&\ds e^{-2\pi i c_{0n} \bar t_n}\ \theta_1(\eta-\bar t_n)^{c_n}\ \prod_{i<n} \theta_1(\bar t_i-\bar t_n)^{c_{in}}\ ,
\eea
each depending on the set of $n$  marked points $\xi$ and $t_i$ and $\eta$ and $\bar t_i$, $i=1,\ldots,n-1$, respectively.  To perform  the relevant complex integral
\be\label{Hochkreuth}
\int_T d^2t_n\ T_n(t_n)\ \overline{T_n(t_n)}
\ee
we proceed as in the previous Subsection \ref{OneP} and regard the integrand of \req{Hochkreuth}
as analytic function in $\Im t_n$, with $t_n=\Re t_n+i\Im t_n$. 
In the complex $\Im t_n$--plane there are $n$ pairs of branch points.
One pair at $\Im t_n=-i(\xi-\Re t_n)$ and $\Im t_n=i(\eta-\Re t_n)$ with phase $e^{\pi i c_{n}}$ and $n-1$ pairs at $\Im t_n=-i(t_l-\Re t_n)$ and $\Im t_n=i(\bar t_l-\Re t_n),\ l=1,\ldots,n-1$ with corresponding phases $e^{\pi i c_{ln}}$. In the complex $\Im t_n$--plane the locations of these branch points is similar than to the case studied before and depicted in Fig.~\ref{Complex_Torus}. Only the pair of points $\Im t_n=-i(\xi-\Re t_n)$ and $\Im t_n=i(\eta-\Re t_n)$ is not complex conjugate to each other and the latter are already aligned along the imaginary axis of $\Im t_n$.
Nevertheless, we may apply the previous steps and consider two cycles $C_\pm$ with $C_+$ connecting all the $n$ branch points $\Im t_n=-i(\xi-\Re t_n)$ and $\Im t_n=-i(t_l-\Re t_n)$, while $C_-$ passing through the $n$ points $\Im t_n=i(\eta-\Re t_n)$ and $\Im t_n=i(\bar t_l-\Re t_n)$.
For a given $\Re t_n\in(-\infty, \infty)$ we introduce the two independent new complex coordinates 
\be\label{changecoordn}
\ba{lcl}
\xi_n&=&\Re t_n+i\Im t_n\ ,\\
\eta_n&=&\Re t_n-i\Im t_n\ ,
\ea\ee
which become real along the imaginary axis of $\Im t_n$, i.e. for $\Im t_n\in i\IR$.
Again, according to \req{changecoordn} the cycles $C_+$ and $C_-$ are combined with 
the cycle describing the $\Re t_n$--integration giving rise to two independent integration paths 
$C_{\xi_n}$ and $C_{\eta_n}$ in the $\Im t_n$--plane, with $C_{\xi_n}$ probing the $n$ zeros of $\theta_1(\xi-\xi_n)$ and $\theta_1(t_l-\xi_n)$ and $C_{\eta_n}$ hitting the zeros of $\theta_1(\eta-\eta_n)$ and $\theta_1(\bar t_l-\eta_n)$, respectively. The $\xi_n$ and $\eta_n$ integrations along the two cycles $C_{\xi_n}$ and $C_{\eta_n}$ are subject to a phase factor 
$\Pi(\eta_n,\xi_n)$ rendering the correct branch cut structure and a splitting function $\Psi(\eta_n,\xi_n)$ carrying the information about  the change of coordinates  \req{changecoordn}. The path $C_{\xi_n}$ may be described by $n-1$ cycles  $\gamma_{k,k+1}$ connecting the $n-1$ remaining  pairs of (unintegrated) points $t_k, t_{k+1}$, with $k=1,\ldots,n-1$ and similarly for $C_{\eta_n}$. As above intersection numbers \req{INTTorus}  can be used to specify  the phase $\Pi(\eta_n,\xi_n)$.
Since the intersection numbers \req{INTERSECTION} are computed locally at points $\nu_{ij}\in \Delta_i\cap\Box_i$ the intersection properties of the cycles $C_\xi,C_\eta$ and  $C_{\xi_n},C_{\eta_n}$ can be computed independently and give rise to a product structure:
\be\label{Osterhofen}
\Pi(\xi,\eta)\ \Pi(\eta_n,\xi_n)\simeq \vev{\gamma_{i,i+1}\otimes KN_z|\gamma_{j,j+1}^\vee\otimes 
  {\overline{KN}}_{\bar z}}\ \vev{\gamma_{k,k+1}\otimes KN_{t_n}|\gamma_{l,l+1}^\vee\otimes 
  {\overline{KN}}_{\bar t_n}}\ .
\ee
Here, $KN_{t_n}$ and ${\overline{KN}}_{\bar t_n}$ describe the relevant local systems 
\bea\label{Nogg}
 KN_{t_n}&=&\ds e^{2\pi i c_{0n} t_n}\ \theta_1(z-t_n)^{c_n}\;\prod_{i=1}^{n-1}\theta_1(t_i-t_n)^{c_{in}}\ ,\\ 
{\overline{KN}}_{\bar t_n}&=&\ds e^{-2\pi i c_{0n} \bar t_n}\ \theta_1(\bar z-\bar t_n)^{c_n}\ \prod_{i=1}^{n-1} \theta_1(\bar t_i-\bar t_n)^{c_{in}}\ ,
\eea
anticipated from \req{Siglalm} with $z,\bar z$ accounting for the (real) parameters $\xi,\eta$, respectively.
Since these local systems assume the form of the integrand \req{Wendelstein} of the Riemann--Wirtinger  integral we can borrow all intersection 
results from Section \ref{RWIsection}, subject to the identifications $c_0\simeq c_{0n}$ and $c_i\simeq c_{in},\;i=1,\ldots,n-1$. 

The above steps are successively  applied to all coordinates $t_k,\bar t_k$, $k<n$ with 
the holomorphic functions
\bea\label{Schweinsberg}
T_k(t_k)&=&\ds e^{2\pi i c_{0k} t_k}\ \prod_{j>k}^{n+1}\theta_1(\xi_j-t_k)^{c_{jk}}\;\prod_{i<k}\theta_1(t_i-t_k)^{c_{ik}}\ ,\\ 
\overline{T_k(t_k)}&=&\ds e^{-2\pi i c_{0k} \bar t_k}\ \prod_{j>k}^{n+1}\theta_1(\eta_j-\bar t_k)^{c_{jk}}\ \prod_{i<k} \theta_1(\bar t_i-\bar t_k)^{c_{ik}}\ ,
\eea
each depending on the set of $n$  marked points $\cup_{j>k}\xi_j$,\;$\cup_{i<k}t_i$ and $\cup_{j>k}\eta_j$,\;$\cup_{i<k}\bar t_i$, respectively, with $\xi_{n+1}=\xi,\eta_{n+1}=\eta$ and $c_{n+1,j}=c_{j}$. The corresponding local systems can  be cast into the forms \req{L1} and \req{L2}, which provide the underlying intersection numbers.
This way all $n$ complex coordinates $z,t_2,\ldots,t_n$ are integrated by mapping the respective  integrand to \req{Startn1}, subject  to the fixing $t_1=0$. 
Again, each complex $t_k$--integration is split into two real integrations $\xi_k,\eta_k$ subject to a phase factor of the form \req{Fully} accounting for the effects from the intersection properties of cycles $\Pi(\xi_k,\eta_k)$ and from the splitting process $\Psi(\xi_k,\eta_k)$. 

The intersection properties are described locally for each sector $k$ by the twisted intersection number \req{INTTorus} and give rise to a product structure as \req{Osterhofen}.
In total, with $\si,\rho\in S_n$ denoting 
the ordering of $n+1$ points we obtain the intersection matrix, cf. \req{REL}
\be\label{Lacherspitz}
\Pi_{\si\rho}=\vev{\Delta(\sigma)\otimes KN|\Delta(\rho)^\vee\otimes 
  \overline{KN}}\simeq e^{i\Phi(\si,\rho)}
\ee
with the pairing
\be\label{Schweinsberg}
|\Delta(\rho)\otimes KN\rng=\int_{\Delta(\rho)} KN\ ,
\ee
referring to the cycle (with $z_1,\bar z_1\equiv 0$ and $z_{n+1}=z,\bar z_{n+1}=\bar z$)
\be\label{Kesselwand}
\Delta(\sigma)=\{(z_2,\ldots,z_{n},z)\in (0,1)\ |\ 0<z_{\si(2)}<\ldots<z_{\si(n)}<z_{\si(n+1)}\}\ ,
\ee
 and the local systems extracted from the integrand of \req{ConsiderN}:
\bea\label{GC2}
KN&=&\ds e^{2\pi i  c_{0} z}\prod_{l=1}^n e^{2\pi i  c_{0l} z_l}\ \theta_1(z-z_l)^{c_{l}}\prod_{i<j}^n\theta_1(z_i-z_j)^{c_{ij}}\ ,\\
\overline{KN}&=&\ds e^{-2\pi i  c_{0} \bar z}\prod_{l=1}^n e^{-2\pi i  c_{0l} \bar z_l}\ \theta_1(\bar z-\bar z_l)^{c_{l}}\prod_{i<j}^n\theta_1(\bar z_i-\bar z_j)^{c_{ij}}\ .
\eea
On the other hand, for each coordinate change \req{changecoordn} the total splitting function
$\Psi$ is composed by a product of the individual functions $\Psi(\xi_k,\eta_k)$, which are  explicitly given in \req{SPLIT}.

Eventually, similar to \req{LAX} altogether we obtain for \req{ConsiderN}:
\bea\label{KITP}
M&=&\ds\sum^\prime_{\si,\rho\in S_n}\vev{\Delta(\sigma)\otimes KN|\Delta(\rho)^\vee\otimes 
  \overline{KN}}  \\
&\times&\ds \int_{\Delta(\sigma)}\lf(\prod_{r=2}^{n+1} dz_r\ri)  KN\int_{\Delta(\rho)} \lf(\prod_{r=2}^{n+1} d\bar z_r\ri)\overline{KN}\ \prod_{k=2}^{n+1}\ \Psi(z_k,\bar z_k)\ .
\eea
Note, that at tree--level the splitting of the complex integrations into real integrations can  conveniently  be described  by inserting the identity \req{Saguaro}.
The same procedure is not applicable  at one--loop due to the change of coordinates \req{changecoordn} resulting in the additional  splitting function 
$\Psi$. The latter is explicitly given by \req{SPLIT}. Besides, in \req{KITP} as a consequence of slicing the torus and introducing the new coordinates \req{changecoordn} for the choice  \req{CHOICE} the splitting function \req{SPLIT} is only inert under shifts in the loop momentum  $\ell\ra\ell+q_n$ or likewise $B$--cycle shifts of $t_{n}$ if $c_n=\h\ap q_nq_{n+1}\in\IZ$. A related issue appears in \cite{Bhardwaj:2023vvm}, where it is suggested to solve this problem by taking an integer parameter $c_n$.

\subsection{One--loop closed string amplitude on the torus}\label{KLTDez}

In the following we shall apply  our results to the one--loop $n$--point closed string torus amplitude. After chiral splitting and introducing a loop momentum $\ell$ this amplitude  reads 
\be\label{Lacherspitze}
\Mc_{n;1}^{closed}(q_1,\ldots,q_n)=\h\;g_{c}^n\;\delta^{(d)}\lf(\sum_{r=1}^{n}q_r\ri)\int_{\Fc_1} \fc{d^2\tau}{\tau_2}\ 
{M}_{n;1}^{closed}(q_1,\ldots,q_n)\ ,
\ee
involving a complex structure integral over the fundamental region $\Fc_1$ of the torus and the integrand
\begin{align}
{M}_{n;1}^{closed}(q_1,\ldots,q_n)&=V_{CKG}^{-1}\int_{-\infty}^\infty d^d\ell\; e^{-\pi \ap\tau_2\ell^2}\;
e^{-\pi i\ap\ell \sum\limits_{i=1}^n q_i(z_i-\bar z_i)}\label{kltone}\\
&\times\lf(\prod_{r=1}^n\int_{E_\tau} d^2z_r\ri)\;\prod_{i<j}|\theta_1(z_i-z_j;\tau)|^{\ap q_iq_j}\ 
\theta_1(z_i-z_j;\tau)^{n_{ij}}\; \overline{\theta_1( z_i- z_j;\tau)}^{\tilde n_{ij}},\nonumber
\end{align}
with some integers $n_{ij},\tilde n_{ij}\in\IZ$. The amplitude \req{Lacherspitze} describes the scattering of $n$ closed string states of  external momenta $q_i\in\IR$ subject to momentum conservation $\sum_{i=1}^nq_i=0$.  The $n$ coordinates $z_r$ represent the vertex operator positions and are integrated over the  torus, i.e. $z_r\in E_\tau$. 
By using contour deformation and Cauchy's theorem on the elliptic curve the $n-1$ complex integrations $z_r$ can be converted to $n-1$ pairs $\xi_r,\eta_r$ of real integrations. This way \req{kltone} can be written in factorized form comprising two open string sectors \cite{Stieberger:2022lss}.

Here, we shall demonstrate the factorization of the $n$ complex integrations in \req{kltone} by applying the  result \req{KITP} and link \req{kltone} to the twisted (co)homology of the Riemann--Wirtinger integral \req{Wendelstein}. 
The discussion on monodromies and analytic continuation does not depend on the specific values of the integers $n_{ij},\tilde n_{ij}$. Therefore, in the following  we shall assume  $n_{ij},\tilde n_{ij}=0$. Furthermore, we restrict to  massless external states, i.e.  $q_i^2=0$. 
We borrow the result \req{KITP} and match it to \req{kltone} by  performing  the following replacements  in \req{KITP},  cf. also \req{CHOICE}:
 \bea\label{Matching}
 \ds c_{ij}&\ra&\ds\h\ap q_iq_j\ \ \ ,\ \ \ c_{0i}\ra-\h\ap \ell q_i\ ,\ i,j=1,\ldots,n\ ,\\[5mm]
 c_k&\ra&\ds\h\ap q_{n+1}q_k\ \ \ ,\ \ \ c_0\ra -\h\ap \ell q_{n+1}\ ,\  k=1,\ldots,n\ ,
\eea
and subsequently:
\be\label{subset}
n\ra n-1\ .
\ee
With these replacements we can write \req{kltone} in the following factorized form
\begin{align}
{M}_{n;1}^{closed}(q_1,\ldots,q_n)&=\int_{-\infty}^\infty d^d\ell\ e^{-\pi\alpha' \tau_2\ell^2}\sum^\prime_{\si,\rho\in S_{n}}\vev{\Delta(\sigma)\otimes KN|\Delta(\rho)^\vee\otimes 
  \overline{KN}}\label{submitted} \\
&\times\int_{\Delta(\sigma)}\lf(\prod_{r=2}^{n+1} dz_r\ri)  KN\int_{\Delta(\rho)} \lf(\prod_{r=2}^{n+1} d\bar z_r\ri)\overline{KN}\ \prod_{k=2}^{n+1}\ \Psi(z_k,\bar z_k)\Bigg|_{{\rm Eq.\ } \req{Matching}\atop \&{\rm Eq.\ } \req{subset}}\ ,
\nonumber
\end{align}
with the choice
\be
z_1=\bar z_1=1\ ,
\ee
the cycles \req{Kesselwand}, the Koba--Nielsen factors \req{GC2} and the intersection matrix \req{Lacherspitz} -- all subject to \req{Matching}.
Note, that by construction the latter can be represented 
as a product of intersection numbers on the one--dimensional space $E_\tau$ computed in 
Section~\ref{RWIsection}.

\sect{Concluding remarks}
\label{Conclusion}

In \req{ContourKLTone} and \req{LAX} for single complex integration with $n$ marked points $t_r\in\IC$ we have formulated one--loop KLT relations in terms of twisted (co)homology. More precisely, we  use intersection numbers \req{INTTorus} of twisted cycles on the elliptic curve, albeit restricting the complex structure modulus to $\Re(\tau)=0$. Furthermore, $B$--cycle monodromy  is supposed to be  cancelled by a  loop momentum integral and the choice \req{CHOICE}.
These intersection numbers are related to the twisted (co)homology of the Riemann--Wirtinger integral \req{Wendelstein}. The decomposition \req{Splitting} w.r.t.~the basis of $n$ twisted cycles  \req{CyclesN} uses twisted intersection numbers, which  refer to the local systems \req{L1} and \req{L2}  with generic (real) kinematic invariants $c_i$ and without constraining the parameter $c_\infty$.  In fact, the latter does not enter in the  relevant intersection data \req{INTTorus} appearing in the final result \req{LAX}.

In contrast, in \cite{Bhardwaj:2023vvm} the double copy relation for single complex integration is conjectured to arise from replacing dual (co)cycles by complex conjugated (co)cycles in the Riemann bilinear relations
and cancelling the multi--validness of the integrand appearing in the definition of  the intersection number. For this description the $B$--cycle monodromy is  closed without a loop momentum integral, but by constraining  $\Im c_\infty$ as in \req{CONSTR}.
This constraint imposes a relation between the complex structure modulus $\tau$, the remaining $n$ unintegrated positions $t_i$ and the parameters $c_i$. For real $c_i$ this becomes~\cite{ghazouani2016moduli}:
\be\label{Pokraka}
\Im c_\infty=c_0\;\tau_2+\sum_{i=1}^nc_i\;\Im t_i\overset{!}{=}0\ .
\ee
At present it is not  clear to us how this constraint subjects the remaining $n$ position integrations $t_i$ or likewise what additional constraints may arise \cite{Bhardwaj:2023vvm}. On the other hand, our approach neither builds up on a variant of Riemann bilinear relations nor it requires \req{Pokraka}, but instead  it is based on  analytic continuation of the torus coordinates subject to: 
\be\label{payoff}
\Re\tau=0\ .
\ee
The requirement \req{Pokraka} may stem from a constraint slicing the torus.  Interestingly, for the assignment \req{CHOICE} the condition \req{Pokraka} is met for 
$\ap\ell q_{n+1}=0$ and $\Im t_i=a$ with a generic real parameter $a\in\IR$, i.e. if (after analytic continuation) all $n$ coordinates $t_i$ are eventually aligned along a $A$--cycle 
 line \cite{Stieberger:2022lss}. See \cite{Stieberger:2023nol} for a working example with the 
 constraint $\ap\ell q=0$.

The generalization to $n+1$ complex integrations \req{ConsiderN} is exhibited in \req{KITP} and  \req{submitted}. The twisted intersection numbers \req{Lacherspitz} are computed from the data 
\req{Schweinsberg}--\req{GC2} and can be mapped for each coordinate $z_i$ to the single complex integration case \req{INTTorus} with generic (real) kinematic invariants $c_i$ and without constraining the parameter $c_\infty$.
From the mathematical side the combinatorial and topological relations \req{KITP} and  \req{submitted} establish a variant of twisted Riemann period relation at genus one -- subject to loop momentum integration. Likewise, our approach is suited to understand how to apply the Riemann bilinear relations for complex conjugated (co)cycles.
However,  the splitting function $\Psi$ prevents from 
describing  the  holomorphic splitting of the $n+1$ complex torus integrations  in lines of the tree--level KLT relations \req{Saguaro} by simply inserting 
\be
\sum^\prime_{\si,\rho\in S_{n}}e^{i\Phi(\si,\rho)}\;|\Delta(\si)\otimes \overline{KN}\rng\ \lng\Delta(\rho)\otimes KN|\ ,
\ee
referring to the data  \req{Schweinsberg}--\req{GC2}. Note also the comment about further possible restrictions at the end of Subsection \ref{GENn}.

The intersection numbers \req{INTTorus} used for our double copy description \req{ContourKLTone} are computed on the torus $E_\tau$. On the other hand, these intersection numbers  do not account for the splitting function \req{SPLIT} describing the cutting of the torus into a cylinder. We believe that intersection numbers \req{INTERSECT} and consequently \req{Lacherspitz} should be defined on a cylinder surface in order to fully account for the cutting procedure and the expression \req{Fully}, cf. also \cite{Stieberger:2021daa}. We will leave this task to our future work.

Finally, it would be very interesting to relax the condition  \req{payoff} and generalize our findings to generic complex structure modulus $\tau$ by means of analytic continuation in the complex $\tau$--plane.

\section*{Acknowledgment}
We thank Johannes Broedel for many insightful  discussions and Oliver Schlotterer for one interesting discussion. This research was supported in part by grant NSF PHY-2309135 to the Kavli Institute for Theoretical Physics (KITP).
\newpage

\appendix 
\section{Direct computation of intersection numbers}
\label{AppA}
\renewcommand\thefigure{\arabic{figure}} 

In this appendix we give a direct and detailed computation of some of the intersection numbers which we have used in this work, namely $I_h([\gamma_{12}],[\gamma_{n,0}^\vee])$, $I_h([\gamma_{1n}],[\gamma_{12}^\vee])$ and $I_h([\gamma_{1n}],[\gamma_{n-1,n}^\vee])$. In order to compute these intersection numbers we need to regularize one of the cycles. By convention and according to \req{Rogers} we choose to regularize cycles $\gamma$ of the  main homology i.e. $\gamma \,\, \in H_1(X,\mathcal{L}_{\omega})$.

\begin{figure}[H]
   \centering
   \includegraphics[scale=.4]{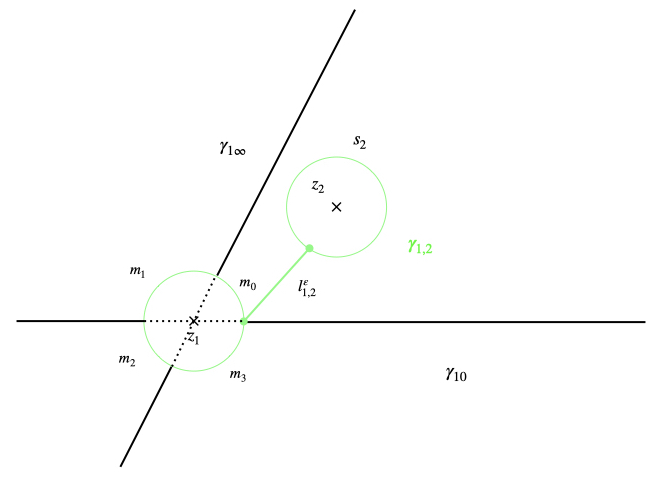}
    \caption{Regularized path $[\gamma_{12}]$ intersecting with $\gamma_{n,0}$}
    \label{regpathcal}
\end{figure}

\begin{figure}[H]
    \centering
    \includegraphics[scale=.35]{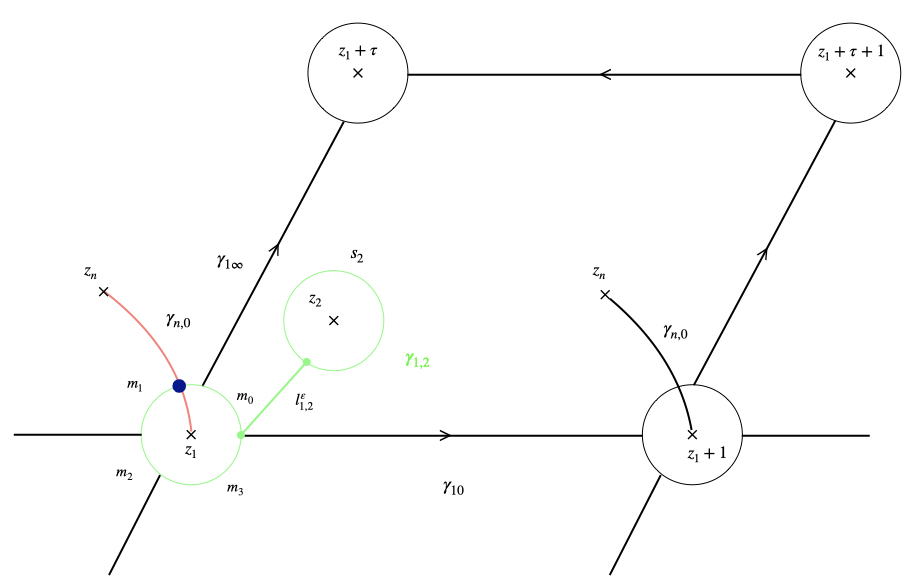}
    \caption{Regularized path $[\gamma_{12}]$ intersecting with $\gamma_{n,0}$}
    \label{regpathcal2}
\end{figure}

We start with the intersection number $I_h([\gamma_{12}],[\gamma_{n,0}^\vee])$. Following the standard convention we regularize the cycle $\gamma_{12}$. The important part of the path regularization is the circle around $z_1$, which we need to decompose into four arcs $(m_0,m_1,m_2,m_3)$ as depicted in Fig. \ref{regpathcal}.
\noindent For each arc we have the local branches defined in the following way 
\begin{equation}
\begin{aligned}
 m_0&=I_0 \otimes T^0(z),\qquad \qquad  &I_0=&[0,\theta], \\
 m_1&=I_1\otimes  (e^{-2\pi i c_0 }T^1(z)),  \qquad \qquad &I_1=&[\theta,\pi],\\
 m_2&=I_2\otimes  (e^{-2\pi i (c_0+c_\infty) } T^2(z)), \qquad \qquad &I_2=&[\pi,\pi+\theta], \\
 m_3&=I_3 \otimes (e^{-2\pi i c_\infty } T^3(z)), \qquad \qquad &I_3=&[\pi+\theta,2\pi], \\
 l_{ij}^\varepsilon&= I_{ij}^\varepsilon \otimes T(z)\Big|_{I_{ij}^\varepsilon}, \qquad \qquad &I_{ij}^\varepsilon=&[z^\varepsilon_i,z^\varepsilon_j],\\
 s_k&=\frac{e^{2 \pi i c_k }}{e^{2 \pi i c_k -1}},
\end{aligned}
\end{equation}
where $z^\varepsilon_i$ corresponds to the circle of radius $\varepsilon$ around $z_i$ and $s_i$ is the monodromy around the $z_i$. With this information, we calculate the intersection number of the regularized path $\gamma_{12}$ with $\gamma_{n,0}^\vee$ as shown in Fig. \ref{regpathcal2}. 
\noindent Obviously, there is only one intersection point which goes through the curve $m_1$. Therefore, we calculate the intersection number at this point and obtain the following expression for this intersection number 
\begin{equation}
\begin{aligned}
 I_h([\gamma_{12}],[\gamma_{n,0}^\vee])&=\Big(\frac{m_0+\rho_1(m_1+m_2+m_3)}{d_1}+l_{1,2}^\varepsilon-\frac{s_2}{d_2}\Big) \cdot \gamma_{n,0}^\vee\\
&=\Big(\frac{m_0+\rho_1(m_1+m_2+m_3)}{d_1} \cdot \gamma_{n,0}^\vee\Big)=\frac{\rho_1(m_1 \cdot \gamma^\vee_{n,0})}{d_1}=\frac{\rho_1 \rho^{-1}_0}{d_1}\ ,
\end{aligned}
\end{equation}
where we have used the notations 
$\rho_i=e^{2 \pi i c_i}$, $d_i=\rho_i-1$.
Therefore we have 
\begin{equation}
\begin{aligned}
& I_h([\gamma_{12}],[\gamma_{n,0}^\vee])=e^{-2 \pi i c_0}\frac{e^{2 \pi i c_1} }{d_1}\ ,
\end{aligned}
\end{equation}
which agrees with \req{Int12n0}.

\begin{figure}[H]
    \centering
    \includegraphics[scale=.35]{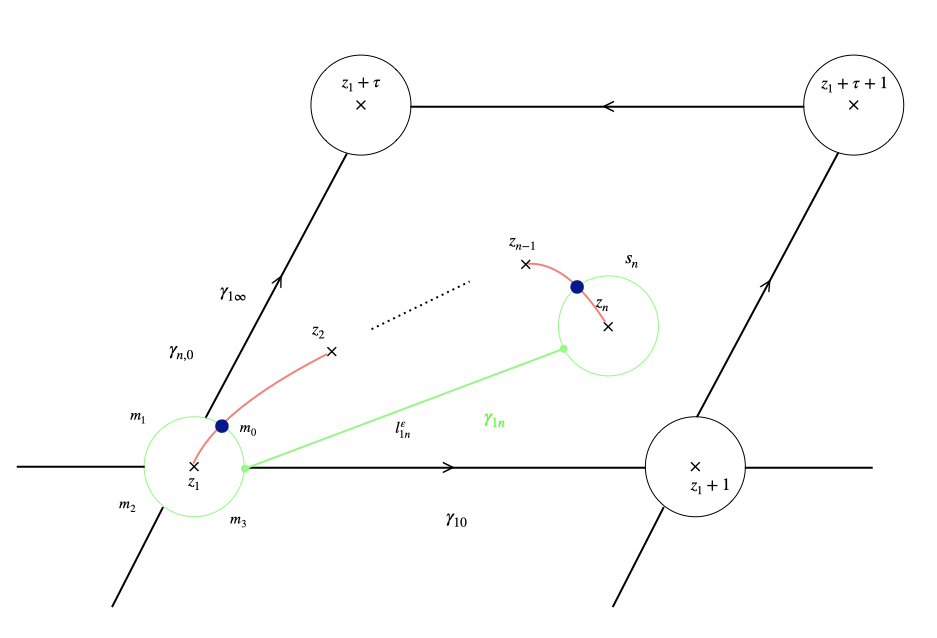}
    \caption{Regularized path $[\gamma_{1n}]$ intersecting with $\gamma_{12}$ and$\gamma_{n-1,n}$ }
    \label{regpathcal3}
\end{figure}

As another important example, we look at the intersection numbers (\ref{1n1}) and (\ref{1n2}) and we demonstrate how we can calculate them directly. For the intersection number (\ref{1n1}), by using standard convention we regularize the cycle associated to $\gamma_{1n}$. We have the intersection diagram, depicted in Fig. \ref{regpathcal3}, where we have indicated the intersection points with blue dots. Note that one can always deform the cycle $\gamma_{1n}$ in such a way that other cycles $\gamma_{i,i+1}$ do not intersect with it. 
We can now compute the two intersection numbers as
\begin{equation}
\begin{aligned}
 I_h([\gamma_{1n}],[\gamma_{12}^\vee])&=\Big(\frac{m_0+\rho_1(m_1+m_2+m_3)}{d_1}+l_{1n}^\varepsilon-\frac{s_n}{d_n}\Big) \cdot \gamma_{12}^\vee\\
 &=\Big(\frac{m_0+\rho_1(m_1+m_2+m_3)}{d_1} \cdot \gamma_{12}^\vee\Big)=\frac{(m_0 \cdot \gamma^\vee_{12})}{d_1}=-\frac{1}{d_1}\\
 I_h([\gamma_{1n}],[\gamma_{n-1,n}^\vee])&=\Big(\frac{m_0+\rho_1(m_1+m_2+m_3)}{d_1}+l_{1n}^\varepsilon-\frac{s_n}{d_n}\Big) \cdot \gamma_{n-1,n}^\vee\\
&=-\Big(\frac{s_n}{d_n} \cdot \gamma_{n-1,n}^\vee\Big)=-\frac{(s_n \cdot \gamma^\vee_{n-1,n})}{d_n}=-\frac{c_n}{d_n} \ ,
\end{aligned}
\end{equation}
where the last minus sign in the first intersection number is due to the topological intersection orientation. 

\begin{figure}[H]
    \centering
   \includegraphics[scale=.35]{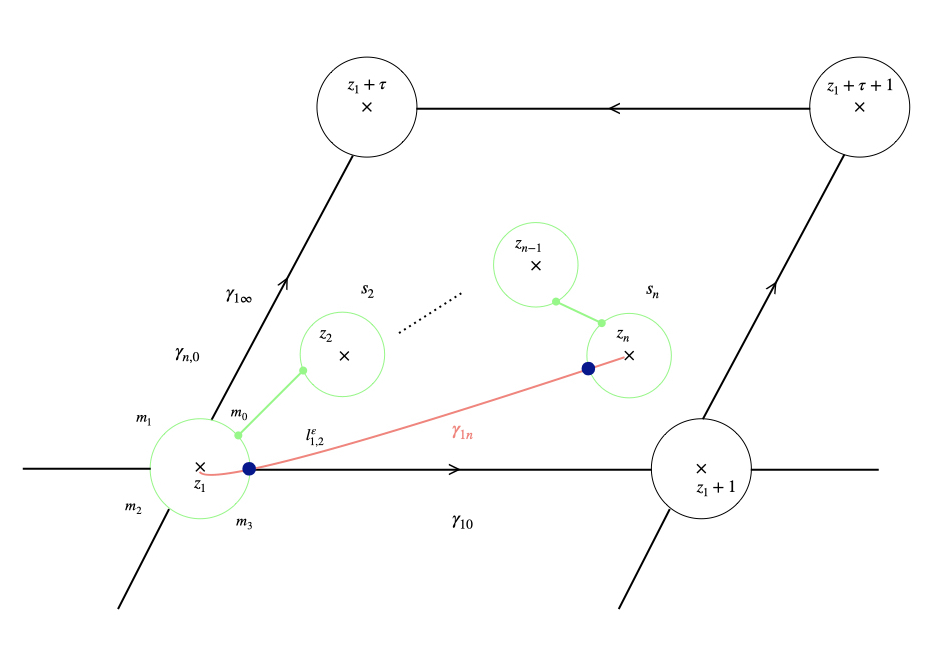}
    \caption{Regularized path $[\gamma_{12}]$ and $[\gamma_{n-1,n}]$intersecting with $\gamma_{1n}$}
    \label{regpathcal1}
\end{figure}

Finally, looking at the other intersection number (\ref{1n2}) and using the standard convention we regularize the paths $\gamma_{i,i+1}$ as depicted in Fig. \ref{regpathcal1}.
\noindent Similar to the last case we can see that we have two intersection points. However, given the locations of the cycles w.r.t.~each other, we have a phase difference relative to the previous case. With this information we obtain the following results: 
\begin{equation}
\begin{aligned}
 I_h([\gamma_{12}],[\gamma_{1n}^\vee])&=\Big(\frac{m_0+\rho_1(m_1+m_2+m_3)}{d_1}+l_{1,2}^\varepsilon-\frac{s_2}{d_2}\Big) \cdot \gamma_{1n}^\vee\\
&=\Big(\frac{m_0+\rho_1(m_1+m_2+m_3)}{d_1} \cdot \gamma_{1n}^\vee\Big)=\frac{(m_0 \cdot \gamma^\vee_{1n})}{d_1}=-\frac{c_1}{d_1}\ ,\\
I_h([\gamma_{n-1,n}],[\gamma_{1n}^\vee])&=\Big(\frac{s_{n-1}}{d_{n-1}}+l_{1n}^\varepsilon-\frac{s_n}{d_n}\Big) \cdot \gamma_{1n}^\vee\\
&=-\Big(\frac{s_n}{d_n} \cdot \gamma_{1n}^\vee\Big)=-\frac{(s_n \cdot \gamma^\vee_{1n})}{d_n}=-\frac{1}{d_n}\ .
\end{aligned}
\end{equation}

\newpage
\addcontentsline{toc}{section}{References}

\bibliography{bibcf}

\end{document}